\documentclass[acmtog]{acmart}

\usepackage{booktabs} 
\usepackage{graphicx}
\usepackage{mathrsfs}

\renewcommand\footnotetextcopyrightpermission[1]{} 

\def\figurePath{}
\def\myfigure#1#2#3{\begin{figure}[t]\centering\includegraphics[width = \linewidth]{\figurePath#2}\caption{#3}\label{fig:#1}\end{figure}}
\def\mycfigure#1#2#3{\begin{figure*}[t]\centering\includegraphics*[clip, width = \linewidth]{\figurePath#2}\caption{#3}\label{fig:#1}\end{figure*}}

\newcommand{\R}{\mathrm{I\!R}}

\def\mysection#1#2{\section{#1}\label{sec:#2}}

\def\bi{\mathbf{i}}
\def\bo{\mathbf{o}}
\def\bh{\mathbf{h}}
\def\bn{\mathbf{n}}
\def\bx{\mathbf{x}}

\newcommand{\bp}{\mathcal{P}}
\newcommand{\bs}{\mathbf{s}}

\newcommand{\bu}{\mathbf{u}}

\newcommand{\dd}{\,\mathrm{d}}

\newcommand{\added}[1]{\textcolor{black}{#1}}

\usepackage{algorithm}
\usepackage{algorithmicx}
\usepackage{algpseudocode}
\usepackage{hyperref}
\usepackage{bm}

\setcopyright{none}

\acmDOI{0000001.0000001_2}

\begin{document}

\title[Constant-Cost Spatio-Angular Prefiltering of Glinty Appearance Using Tensor Decomposition]{Constant-Cost Spatio-Angular Prefiltering of Glinty Appearance Using Tensor Decomposition}

\author{Hong Deng}
\affiliation{%
  \institution{School of Computer Science and Engineering, Nanjing University of Science and Technology}
  \streetaddress{200 Xiaolingwei Rd}
  \city{Nanjing}
  \postcode{210094}
  \country{China}}

\author{Yang Liu}
\affiliation{%
  \institution{School of Computer Science and Engineering, Nanjing University of Science and Technology}
  \streetaddress{200 Xiaolingwei Rd}
  \city{Nanjing}
  \postcode{210094}
  \country{China}}

\author{Beibei Wang$^{*}$}
\thanks{Corresponding author.}
\orcid{0000-0001-8943-8364}
\affiliation{%
  \institution{School of Computer Science and Engineering, Nanjing University of Science and Technology}
  \streetaddress{200 Xiaolingwei Rd}
  \city{Nanjing}
  \postcode{210094}
  \country{China}}
\email{beibei.wang@njust.edu.cn}

\author{Jian Yang}
\affiliation{%
  \institution{School of Computer Science and Engineering, Nanjing University of Science and Technology}
  \streetaddress{200 Xiaolingwei Rd}
  \city{Nanjing}
  \postcode{210094}
  \country{China}}
\email{csjyang@njust.edu.cn}

\author{Lei Ma}
\affiliation{%
  \institution{National Engineering Laboratory for Video Technology, Peking University}
  \city{Beijing}
  \country{China}}
\email{lei.ma@pku.edu.cn}

\author{Nicolas Holzschuch}
\affiliation{%
 \institution{University Grenoble Alpes, Inria, CNRS, Grenoble INP, LJK}
 \streetaddress{655, avenue de l'Europe}
 \city{Grenoble}
 \country{France}}
\email{Nicolas.Holzschuch@inria.fr}

\author{Ling\-Qi Yan}
\affiliation{%
  \institution{University of California, Santa Barbara}
  \streetaddress{2119 Harold Frank Hall}
  \city{Santa Barbara, CA, 93106}
  \country{USA}
}
\email{lingqi@cs.ucsb.edu}

\renewcommand\shortauthors{Deng et al.}

\begin{abstract}
The detailed glinty appearance from complex surface microstructures enhances the level of realism, but is both space- and time-consuming to render, especially when viewed from far away (large spatial coverage) and/or illuminated by area lights (large angular coverage). In this paper, we formulate the glinty appearance rendering process as a spatio-angular range query problem of the Normal Distribution Functions (NDFs), and introduce an efficient spatio-angular prefiltering solution to it. We start by exhaustively precomputing all possible NDFs with differently sized positional coverages. Then we compress the precomputed data using tensor rank decomposition, which enables accurate and fast angular range queries. With our spatio-angular prefiltering scheme, we are able to solve both the storage and performance issues at the same time, leading to efficient rendering of glinty appearance with both constant storage and constant performance, regardless of the range of spatio-angular queries. Finally, we demonstrate that our method easily applies to practical rendering applications that were traditionally considered difficult. For example, efficient bidirectional reflection distribution function (BRDF) evaluation accurate NDF importance sampling, fast global illumination between glinty objects, high-frequency preserving rendering with environment lighting, and tile-based synthesis of glinty appearance.

\end{abstract}


%
%


%
%

\begin{teaserfigure}
  \centering
   \includegraphics[width=\textwidth]{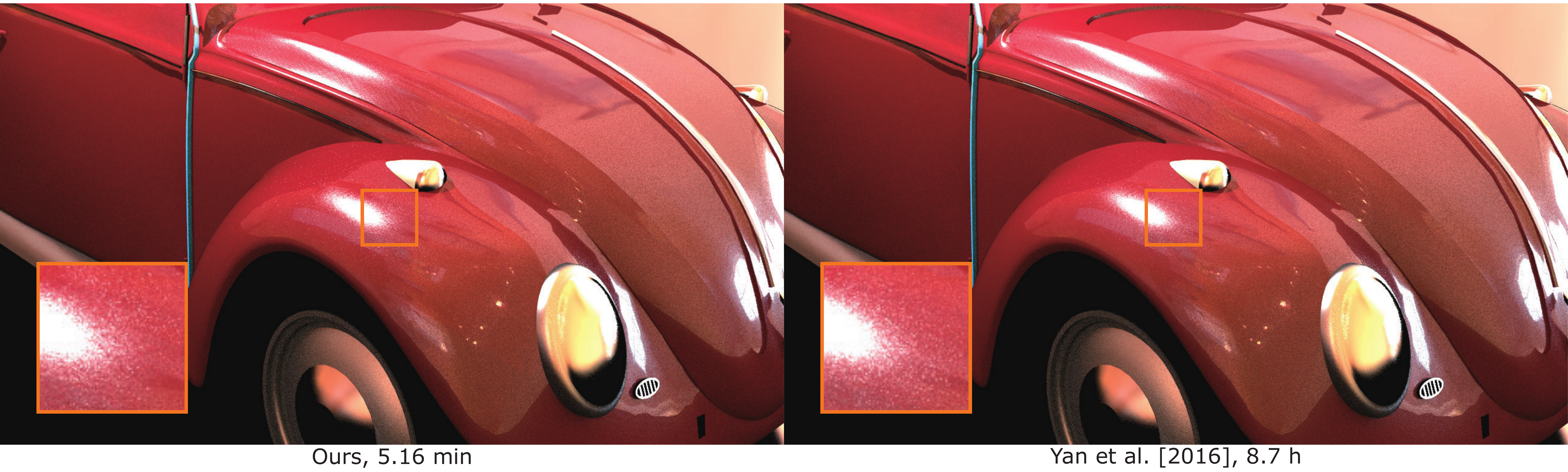}
  \caption{Comparison between our method and Yan et al.~\shortcite{yan2016position} on the Car scene, using a seamlessly tileable normal map of metallic flakes with resolution $2K \times 2K$ under geometric optics. The rendered glints are almost identical, but our method costs less than $1\%$ of time compared to Yan et al.~\shortcite{yan2016position}, and does not scale with the view distance.}
	\label{fig:teaser}
\end{teaserfigure}

\maketitle

\mysection{Introduction}{Introduction}

Many materials in the real life have microstructures, such as the tiny metallic flakes in car paints, and scratches and brushes on heavily used cutleries. These microstructures have their own way of interaction with the light, and when aggregated, change the overall visual appearance of the materials. Microstructure rendering~\cite{Yan14Rendering} studies the modeling and rendering of such details, and has increased the realism to rendering individual specular highlights (i.e. glints), a core effect in computer graphics. But in order to represent, store and render the complex appearance from details, the storage is usually prohibitively costly and the performance is often too slow, even for offline productions such as animations.

The fundamental difficulty is that in order to bring out the complex light transport from microstructures, one has to define all the details. For general appearance, it is common practice to use high-resolution normal maps or heightfields to explicitly define every microfacet. For example, Yan et al.~\shortcite{yan2018wave} define heightfields with resolutions up to one micron per texel. The size of such high-resolution details is large -- $10K \times 10K$ texels just to cover one square centimeter, and it grows with the area of surfaces.

However, an even bigger issue is the computation. During rendering, it is the NDF of the microstructures that determines specific appearance. However, the more microstructures are being considered, the longer it takes to resolve the interactions between the light and the microstructures. This imposes strong difficulties on the rendering performance. When the viewpoint is far away from the objects, the surface region covered by one pixel, a.k.a. pixel footprint, will include more microstructures inside. Also, when large area lights exist, multiple NDF queries will be needed for different samples on the lights. Noise will emerge, and will consume a lot of samples to diminish, because the microstructures contribute to different parts of the light quite differently. Recent works (Gamboa et al.~\shortcite{Gamboa:2018:GXD}) have solved this issue with Summed Area Table (SAT) or Integral Histogram (IH), at cost of expensive storage or limited microfacet models (Beckmann only). 

From these difficulties, we can see that the performance of rendering glinty appearance heavily depends on the spatial and angular coverages of NDF queries. Therefore, at the core of rendering glinty appearance from microstructures is an efficient method that performs \emph{spatio-angular prefiltering} of NDFs. That is, with some proper treatments before rendering, we should be able to quickly find the NDF value from microfacets within an arbitrary pixel footprint towards an arbitrary range of solid angles. Preferably, the time it takes to answer such questions should be independent of the query sizes, both spatially and angularly, known as constant performance. Even better, if the prefiltering can be combined with dynamic synthesis methods of the microstructure, we will be able to generate microstructures in an infinite positional range, while keeping the prefiltering only on the given finite sized microstructures. This gives us constant storage.

In this paper, we present a precomputation-based prefiltering approach that answers spatio-angular range queries of NDFs, achieving both constant performance and constant storage, a.k.a. constant cost.
We start from an exhaustive precomputation of all NDFs at all possible centers and sizes of pixel footprints, to cover all possible NDF queries during rendering. We organize the precomputed data into a 3D ``NDF tensor'', and use tensor rank decomposition to compress it. Then we propose a novel decompression scheme that answers arbitrarily-sized spatio-angular queries on the original NDF tensor at constant performance.

Our prefiltering scheme leads to efficient rendering of glinty appearance. We demonstrate that our method elegantly applies to practical rendering operations/applications that were traditionally considered difficult: accurate NDF importance sampling, fast global illumination between glinty objects, high-frequency preserving rendering with environment lighting. And we complete our constant-cost goal by combining our prefiltering with tile-based synthesis methods that achieves constant storage of rendering a large range of microstructures.

Compared with previous glinty appearance rendering methods, our method is the first that does not have to tradeoff between quality and cost. In practice, the performance of our method is also constant w.r.t. view distances and light sizes, and is 1-2 orders of magnitude faster than traditional approaches. And the storage cost of our method is constant to the granularity of the scene, taking up only several megabytes (e.g. for $1K\times 1K$ patches) to hundreds of megabytes (e.g. for $8K\times 8K$ patches) after compression.

To summarize, our contributions include: 
 \begin{enumerate}
 	\item \added{a NDF precomputation-based approach with compact compression and decompression using tensor decomposition}, which enables fast evaluation of BRDF with large footprint,
	\item a large range of applications to several difficult rendering problems: accurate NDF importance sampling, multiple-bounce glints rendering and prefiltering, and
 	\item a combination with tile-based texture synthesis methods, resulting in constant storage for glints rendering.
 \end{enumerate}

\section{Related Work}
\label{sec:related}

In this section, we first briefly review previous work on microstructure rendering and synthesis, and then introduce related work on precomputation and prefiltering.

\textbf{Microstructure rendering.} Yan et al.~\shortcite{Yan14Rendering} proposed to simulate the spatially and directionally varying appearance using patch-local normal distribution functions ($\bp$-NDFs), which is later accelerated by Yan et al.~\shortcite{yan2016position} via a position-normal distribution method. Recently, Yan et al.~\shortcite{yan2018wave} managed to wave optics effects. As analyzed in Sec.~\ref{sec:Introduction}, all these methods share a common problem with performance that grows with the size of the pixel footprint.

More generally, Zeltner et al.~\shortcite{Zeltner:2020:Specular} exploit the specular manifold to efficiently find specular paths in rendering complex appearance. Wang et al.~\shortcite{wang:2020:specular} introduce the idea of path cuts to find all specular paths of arbitrary lengths. These methods aim at efficient specific-purpose light transport, possibly with microstructure, but do not focus on appearance models. Our method makes no assumption to light transport methods. It automatically fits the Multiple Importance Sampling (MIS) framework, and naturally handles multiple bounces of light between complex surfaces.

\textbf{Microstructure synthesis.} Considerable efforts have been made to relieve the storage issue in microstructure rendering. And they can be subscribed into two categories. The first category is the procedural creation of the microstructure, realizing the actual distributions of microstructures on the fly from a list of predefined rules \cite{Jakob14Discrete,Wang:2018:Glint,Wang:2020:RealtimeGlint,Chermain:2021:GlintAA,zirr2016real,Raymond:2016:MRS,Werner:2017:Scratch,Velinov:2018:WaveOpticalScratches}. These methods work well for specific effects, such as discrete glitters and scratches, but do not support general appearances. The other category of microstructure synthesis is by-example \cite{Wang:2020:Example,Zhu:2019:svbrdf}, using texture synthesis~\cite{Heitz:2018:HBN, Cohen:2003:WangTiles} or using a Generative Adversarial Network to generate NDF images \cite{Kuznetsov:2019:GlintGan}. Our method combines with the by-example synthesis idea, bringing out the constant storage property.

\textbf{Precomputation} followed by efficient compression has been widely adopted in computer graphics. Precomputed Radiance Transfer (PRT) precomputes complex light transport and compresses them in order to allow real-time rendering with complex lighting~\cite{Sloan:2002:SH,Sloan:2003:CPCA,NG:2004:wavelet,Tsai:2006:SG,Wang:2009:SG,xu:2013:ASG}. Precomputation also takes place pervasively in appearance modeling and rendering. Yan et al.~\shortcite{yan2017furbssrdf} use precomputed 2D profiles to describe the scattering behavior in different types of animal fur fibers. Donner et al.~\shortcite{Donner:2009:EBSSRDF} exhaustively precompute one BSSRDF for every different combination of scattering parameters, and use confocal ellipses to fit the resulting data, leading to an efficient empirical BSSRDF model. Wang et al.~\shortcite{Wang2020Premultiple} propose to represent the multiple scattering in participating media with a precomputed table, which is later improved to neural networks~\shortcite{Ge2019isos} for further compression. Rainer et al.~\shortcite{Rainer2019Neural,Rainer2020Unified} uses a neural network to compress measured 6D Bidirectional Texture Functions (BTFs) to relieve the heavy data storage. We analyze the pros and cons of precomputation in Sec.~\ref{sec:motivation}, and exploit precomputation to achieve constant performance.

\textbf{Prefiltering} can be treated as a specific type of precomputation on the appearance, in order to improve run-time performance. For example, to enable efficient level of detail rendering of cloth, Wu et al.~\shortcite{wu2019accurate} prefilter the heightfield and Zhao et al.~\shortcite{Zhao:2016:Anisotropic} prefilter anisotropic participating media. Prefiltering has also been used in real-time rendering with environment maps \cite{Karis:2013:split} and to account for the curvature of surfaces \cite{Kaplanyan:2016:filtering}.

Specifically, prefiltering has been applied to glinty appearance rendering. Belcour et al.~\shortcite{Belcour:2017:prefiltering} prefilter color-, normal-, and displacement-mapped appearance in the context of multi-bounce global illumination with footprints predicted by covariance tracing. Gamboa et al.~\shortcite{Gamboa:2018:GXD} precompute all possible NDFs as histograms for arbitrary spatial range queries at constant performance. However, the resolution of the NDFs is too low ($9\times 32$) to preserve the glinty appearance (e.g. metallic flakes) angularly. It focuses on fast performance during rendering, but still suffers from prohibitively expensive storage cost in practice. Concurrent work \cite{Atanasov:2021:Multiscale} proposes a normal map filtering approach via inverse binning mapping, throwing microstructures into corresponding directional bins as a preprocess, taking advantage of the fact that directional resolution can be fixed regardless of the sizes of pixel footprints. However, this work is limited to Beckmann function.

 \begin{table}[]
  \caption{\label{tab:symbols}Symbols used in the paper.}
\begin{tabular}{cl}
\toprule
   \multicolumn{1}{c}{symbol}
	 & \multicolumn{1}{l}{definition} \\	
		\midrule
		$\bi, \bo$       & incident and outgoing directions\\ \hline
		$\bh$ & half vector \\ \hline
		$f_r(\bi,\bo)$   & BRDF\\ \hline
		$F(\bi\cdot\bh)$ & Fresnel term\\ \hline
		$G(\bi\cdot\bh)$ & Shadowing-Masking function \\ \hline
		$	D(\bi,\bo)$ & normal distribution function\\ \hline
		$ D_\bp(\bx,\bh)$& $\bp$-NDF\\ \hline
		$D^{*}(\bx,\overline{\psi})$ & wave optics GNDF\\ \hline		
		$G(\bx;\mu,\sigma)$ & Gaussian function\\ \hline	
		$G_c$ & coherence area\\ \hline
		$s$   &sample stride on normal map  \\ \hline 
	 	$t$   & image block size of a NDF image  \\ \hline 
		$L$   &count of non-empty NDF image blocks in a cluster  \\ \hline 
		$R$  & rank of compression \\ \hline
		$\bm{X}_r, \bm{Y}_r$ & tensor vectors with size $R \times t$ \\ \hline
		$\bm{Z}_r$ & tensor vectors  with size $R\times L$ \\ \hline
		$C$ & tensor vectors with size $R$\\ 
	
\bottomrule

\end{tabular}
\end{table}

\section{Preliminaries and motivation}
\label{sec:motivation}

In this section, we first briefly provide the fundamentals of rendering complex appearance. Then we analyze the related storage and performance issues of existing methods to motivate our approach. For clarity, we list the symbols used throughout our paper in Table~\ref{tab:symbols}.

\subsection{Preliminaries}

Traditionally, people use the microfacet model~\cite{Cook1982REFLEC,walter2007mmrt} to describe the BRDF at a point $\bx$ on a surface:
\begin{equation}
	f_r(\bi,\bo)=\frac{F(\bi\cdot\bh)\ G(\bi\cdot\bh)\ D(\bi,\bo)}{4\ (\bi\cdot\bn)\ (\bo\cdot\bn)}, 
	\label{eqn:microfacet}
\end{equation}
where $\bi$, $\bo$ are the incident and outgoing directions and $\bn$ is the normal of the macro surface. $F$ is the Fresnel term giving the total amount of reflection, $G$ is the shadowing-masking term considering self-occlusions from microfacets, and most importantly, $D$ is the NDF term, statistically defining the distribution of microfacets' normals at/around $\bx$.
Yan et al.~\shortcite{Yan14Rendering} and subsequent work replace the statistical distribution $D$ with an actual distribution of normals $D_\bp$ (the $\bp$-NDF) in an area around $\bx$, namely the pixel footprint $\bp$ seen through each pixel. This leads to a $\bp$-NDF defined as
\begin{equation}
	D_\bp(\bx,\bh) = \int_{\R^2} G_p(\bu;\bx,\sigma_p)\ G_r(\bn(\bu);\bh,\sigma_r)\dd \bu, 
	\label{eqn:geom}
\end{equation}
where $G(\bx;\mu,\sigma)$ is a 2D Gaussian function centered at $\mu$ with a standard deviation of $\sigma$. $G_p(\cdot;\cdot,\sigma_p)$ and $G_r(\cdot;\cdot,\sigma_r)$ determine the size of the pixel footprint and the ``intrinsic roughness'' of each microfacet, respectively. $\bh=\frac{\bi+\bo}{\vert \bi + \bo \vert}$ is known as the half vector.

In this way, the actual distributions bring out the variation of appearance at different pixels, resulting in interesting glinting effects, especially when the camera or the light moves.

The wave optics variation of the NDF~\cite{Kuznetsov:2019:GlintGan} takes diffraction into account, producing colored appearance. Strictly speaking, the concept of NDFs do not exist in wave optics. Commonly adopted wave optics theories (Harvey-Shack or Kirchhoff) directly predicts the BRDF using Fourier transform of the spatially-varying phase shift of light induced by a heightfield~\cite{yan2018wave}, in a coherence area \emph{of fixed size} (typically $5-10$ microns as the standard deviation of a 2D Gaussian) where the light interferes in a non-linear fashion.

For convenience, Kuznetsov et al.~\shortcite{Kuznetsov:2019:GlintGan} define a ``wave optics GNDF (generalized NDF)'', formally written as 
\begin{equation}
	D^{*}(\bx,\overline{\psi})=\frac{4}{A_c \lambda^2}\left| 
	\int_{\R^2} G_c(\bs;\bu,\sigma_c) e^{-i\frac{4\pi}{\lambda}\left[h(\bs)+ \overline{\psi} \cdot \bs\right]} \dd\bs
	 \right|^2,
	\label{eqn:wave_cohereance}
\end{equation}
where a heightfield $h$ is required instead of normals, $G_c$ is the coherence area, $\overline{\psi}$ is the first two components of the sum $\frac{\bi + \bo}{2}$, and $A_c$ is a normalization factor. We refer readers to Kuznetsov et al.~\shortcite{Kuznetsov:2019:GlintGan} for more details.

When the pixel footprint is greater than the coherence area, which is almost true at all times even when viewed from closeup, we are able to quickly integrate over the pixel footprint to get the (linearly) aggregated GNDFs as
\begin{equation}
	D_\bp^{*}(\bx,\overline{\psi})=\frac{\int_{\R^2} G_p(\bx;\bu,\sigma_p)\ D^{*}(\bu,\overline{\psi})\dd\bu}{\int_{\R^2} G_p(\bx;\bu,\sigma_p)\dd\bu}. 
	\label{eqn:wave_cohereance}
\end{equation}

As we can see, the NDFs for both geometric optics and wave optics are able to be defined consistently as 2D images with given sizes of pixel footprints. Also note that, as shown in Figure~\ref{fig:precompute}, even when the pixel's footprint becomes rather large, the NDF still contains a lot of high-frequency information. This indicates that the NDF images still needs accurate computation.

\subsection{Motivation}

\myfigure{motivation}{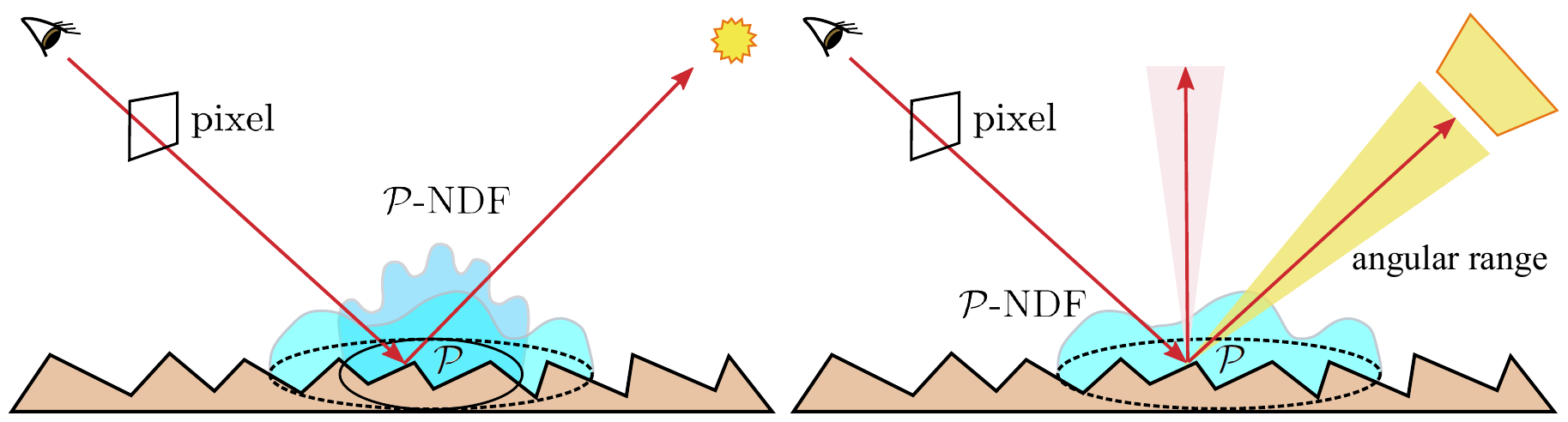}{Motivation of spatio-angular filtering of glinty appearance. Left: larger spatial footprint results in smoother NDF, but takes longer to compute. Right: larger angular range results in smoother NDF, but previously solved with heavy sampling rather than simple range query.}

From the way $\bp$-NDFs and GNDFs are computed (Eqns.~\ref{eqn:geom} and~\ref{eqn:wave_cohereance}), we immediately find that the complexity is corresponding to the spatial size of the pixel footprint. This is undesired and often against intuition. For a large pixel footprint (Fig.~\ref{fig:motivation} (left)), it takes longer to compute or query the NDF, even though there are fewer high frequency details in the NDF. Also, when an area light or environment light is involved (Fig.~\ref{fig:motivation} (right)), sampling the solid angle subtended by the light results in different incident directions, indicating that the NDF needs to be queried multiple times at different half vectors. However, this is equivalent to rendering with a fixed incident direction using an angularly blurred NDF.

The above analysis inspires us to design a spatio-angular prefiltering scheme --- given any spatial range on the surface and any angular range on half vectors, we want to find the corresponding NDF image (spatial filtering), then find the average of NDF values within a block on this image (angular filtering). Even better, we should be able to skip producing the entire NDF image and only focus on the angular block, and we should avoid looping over all places inside this block but directly acquiring the average value. This is crucial to achieve constant performance.

To carry out an actual spatio-angular prefiltering method, we start by considering the following question: is it possible to exhaustively precompute all the NDF images for all possible pixel footprints, until the footprint is large enough so that the NDFs become smooth? In this way, the heavy computation of NDFs will become a simple query of the precomputed data, and no matter how large a pixel footprint is, the query could be performed in constant time. The answer is yes, and we demonstrate the possibility in Sec.~\ref{sec:method}.

However, precomputation-based approaches often have three fundamental problems that prevent their practical use:
\begin{enumerate}
	\item heavy data storage, especially for high-resolution normal maps/heightfields used to define the microstructures, 
	\item difficult partial evaluation, i.e. hard to extract part of the compressed data instead of the entire chunk during rendering, and, 
	\item fixed representation, that is, the precomputed data usually cannot be manipulated and used for synthesis.
\end{enumerate}

We challenge all these limitations from precomputation, and prove that even exhaustive precomputation can be made both elegant and practical. 
We elaborate our method in the next sections, starting by introducing the technical part of our method, explaining how to precompute, compress and decompress data, leading to arbitrary constant-cost spatio-angular prefiltering (Sec.~\ref{sec:method}). With our prefiltering approach, we introduce applications to meet practical rendering needs, such as importance sampling, global illumination and dynamic appearance synthesis (Sec.~\ref{sec:rendering}). Finally, we show high quality results with constant cost and compare with previous work (Sec.~\ref{sec:results}).

Like most other works \cite{Yan14Rendering,yan2016position,Wang:2020:Example} on glinty appearance, we only focus on the NDF term in microfacet models. We leave the (also important) shadowing-masking terms and local multiple bounces among microfacets to the future work (global multiple bounces between objects will be properly addressed). Also, since angular prefiltering is aimed at high performance, similar to other angular prefiltering work~\cite{Gamboa:2018:GXD}, we do not deal with drastic visibility changes in the angular range. Apart from these, no further assumptions are made w.r.t. specific types of microstructure (though quality may vary, as will be analyzed thoroughly), granularity of the scene, optical models (geometric optics or wave optics), and so on.

\section{Spatial-Angular Prefiltering: Precomputation, Compression and Decompression}
\label{sec:method}

In this section, we focus on data preparation. We first explain how to perform exhaustive precomputation. Then we introduce our efficient compression scheme of the precomputed data, together with an accurate way of decompression that enables us to quickly answer the average value within a spatial and angular range.

\subsection{Precomputation}
\label{sec:precompute}

\myfigure{precompute}{ndfmipmap.pdf}{We precompute NDFs at different levels. At each level, NDFs are computed at uniformly sampled locations (the blue dot) with the same footprint size (green transparent circle). Each higher level has $2 \times 2$ sparser sample count and twice larger footprint size than the previous level. The bounding box of the footprint size for each level is $30\times 30$, $60\times 60$ and $120\times 120$ pixels. Normal map: isotropic noise (2K$\times$2K).}

To avoid expensive on the fly computation of NDFs during rendering, we propose to precompute the NDFs at discrete locations on the normal map with different pixel footprint sizes. We arrange such pixel footprints in a pre-determined multi-level structure, as Fig.~\ref{fig:precompute} illustrates. In each level, we sample the pixel footprints' locations (centers) uniformly in a grid, and adjust their sizes accordingly so they can cover the entire normal map. The higher the levels are, the sparser we sample the pixel footprints, and the larger these pixel footprints will be, in order to guarantee full coverage.

For each sampled pixel footprint, we compute an NDF image using the method by Yan et al.~\shortcite{yan2016position}. As a result, our multi-level structure is similar to a texture mipmap. However, note the significant difference: each ``texel'' contains a color value from any level in a texture mipmap, but it is \emph{an entire NDF image} in our case. With our multi-level structure, for any query with an arbitrary pixel footprint (location and size), we are able to find the NDF image immediately via trilinear interpolation between different samples in the same level and between different levels. In this way, we have successfully enabled constant performance spatial filtering which is irrelevant to the footprint size.

\textbf{Practical choices.} We generate the pixel footprint samples in a bottom-up fashion, as shown in Fig.~\ref{fig:precompute}. In the lowest level, we sample one pixel footprint with a fixed stride $s=32$, i.e. drawing one sample every $s\times s$ texels. And the size of each pixel footprint is represented with a 2D Gaussian of standard deviation $\sigma_p=1.5s/\sqrt{12}$, as suggested by Yan et al.~\shortcite{yan2016position}. Starting from the second finest level, the sampled pixel footprints are $2\times2$ sparser and their sizes are $2\times2$ larger than those in the next lower level. We continue this sampling process until the NDF images for the neighboring levels are similar (with small MSE) or the stride reaches the normal map size.

As introduced in the background, one advantage of our precomputation is that it does not depend on specific optical models. For each sampled pixel footprint, we compute its NDF image using Yan et al.~\shortcite{yan2016position} under geometric optics, and Yan et al.~\shortcite{yan2018wave} under wave optics. The only difference is that the NDF images under wave optics are colored. Similar to ~\cite{Kuznetsov:2019:GlintGan}, we compute these NDF images using 8 spectrum samples then convert them to RGB images. 
\subsection{Compression}
\label{sec:compress}

\myfigure{cluster}{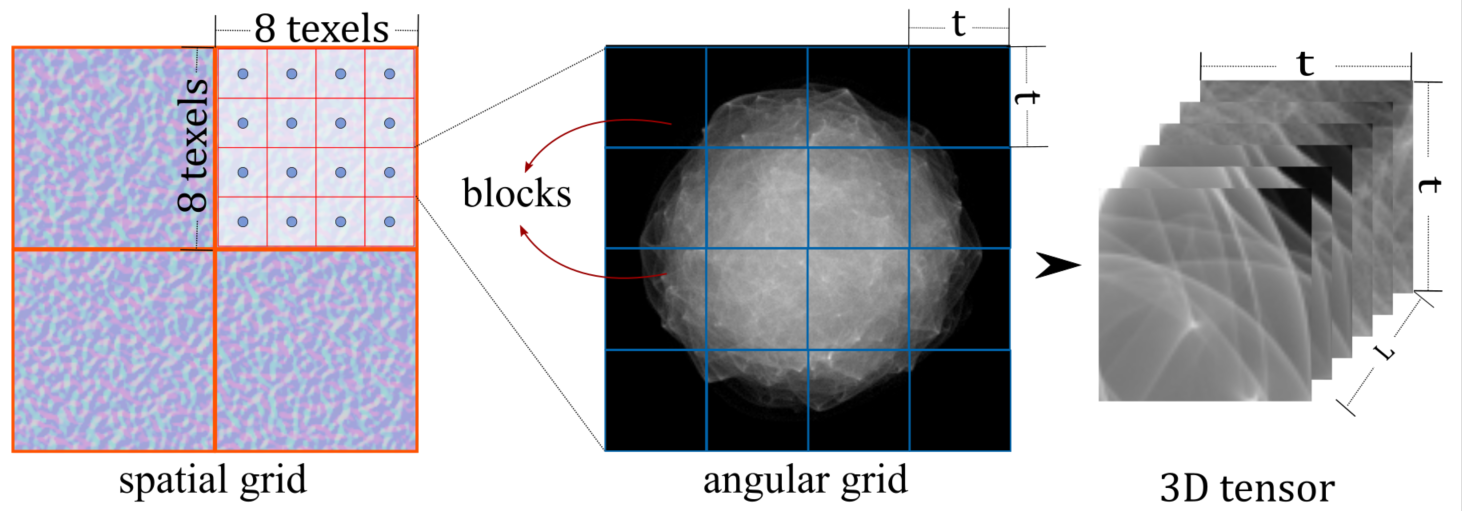}{Before compression, we cluster the NDF image blocks both positionally (on the normal map) and angularly (on the NDF image). All the image blocks which locate in the same angular grid and the same spatial grid form a 3D tensor, and will be compressed together. $L$ is the count of non-empty NDF image blocks in a cluster. }

\myfigure{tcPoint}{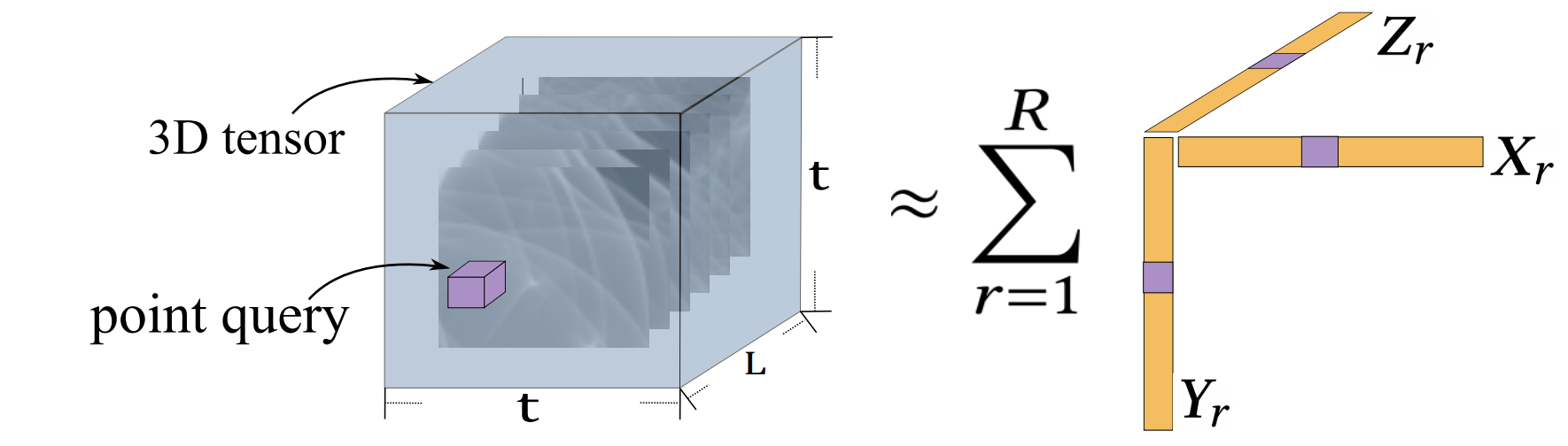}{Illustration of tensor decomposition. Tensor decomposition factorizes a 3D tensor (clustered blocks) into a linear combination of rank one tensors ($\bm{X}_r, \bm{Y}_r$ and $\bm{Z}_r$), where the number of the rank one tensors is $R$ in this figure. With the factorized tensor vectors, point query (purple box) could be performed.}

The precomputed NDF images can be used for BRDF evaluation directly, however, the precomputed data can be storage-consuming, which makes it less practical. We propose an efficient compression scheme of the precomputed NDF images in this section.

An important observation is that the NDF images from higher levels must contain the features from lower levels, therefore we propose an immediate solution that stacks all the NDF images together into a 3D tensor and compresses it using tensor decomposition.

Moreover, we find that there are similar parts (e.g. the high frequency curves) and large unoccupied black regions (especially when the overall appearance is glossy) inside each NDF image, which can be better utilized. Hence, we subdivide each NDF image into blocks in an angular grid (Fig.~\ref{fig:cluster} (middle)).

However, stacking all these image blocks together results in a very ``thin'' tensor, which is not desired for tensor compression. Therefore, we further cluster these image blocks into different groups. Each group contain image blocks generated from pixel footprints with similar locations and levels. And then we compress each group individually using 3D tensor decomposition.

The tensor decomposition generalizes the Singular Value Decomposition (SVD) in 2D into higher dimensions. Specifically, in our case, it factorizes each 3D tensor in a group $j$ into a weighted sum of outer products from $3$ vectors, as shown in Figure~\ref{fig:tcPoint}. Each outer product results in a rank one tensor, and we keep the most significant $R$ ranks with the largest weights. Therefore, we have 
\begin{equation}
	D \approx \hat{D} = \sum_{r = 1}^{R} {C_{r} \otimes \bm{X}_r \otimes \bm{Y}_r \otimes \bm{Z}_r},
\label{eq:compress}
\end{equation}
where $D$ is the original 3D tensor in the group, approximated as $\hat{D}$ after compression, $\bm{X}$, $\bm{Y}$ and $\bm{Z}$ are 1D vectors of length $R\times t$, $R \times t$ and $R \times L$, respectively. And $C$ is the coefficient for different ranks. For simplicity, we ignore the group index $j$ here.

\textbf{Practical choices.} Angularly, we set the size of each image block as $t\times t$, and we choose $t=16$ in practice. Spatially, we do not use optimization to guide how to form clusters\cite{Sloan:2003:CPCA}. Instead, we refer to a simple deterministic method to perform clustering. That is, we divide the entire normal map into a spatial grid, separating the normal map into $8\times 8$ texel regions. We cluster all the NDF images (blocks) into one group as long as their corresponding pixel footprints are centered inside the same region, as shown in Figure~\ref{fig:cluster}.

In each group, we stack the image blocks along the third dimension, resulting in a 3D tensor with size $t \times t \times L$, where $L$ is the number of image blocks. Then we conduct tensor decomposition for each group. Specifically, we perform Canonical Polyadic Decomposition~\cite{Hitchcock:1927:CPD} using alternating least squares with rank set as $R$, with maximum error set as $10^{-4}$ and maximum iteration count set as 500.


We set the maximum rank $R$ as $16$ or $32$, based on specific types of normal maps, as will be elaborated in Sec.~\ref{sec:results}. Compared to the original 3D tensor, we have reached a compression ratio between 0.72\% $\sim$ 3.76\%. This ratio is not constant for different normal maps, because we further optimize our compression scheme by throwing away complete blank image blocks, which is especially efficient for glossy appearances. With our choice of $R$, we have a good balance practically between the compression ratio and the compression quality, which is demonstrated in Fig.~\ref{fig:rank_curve}. Regarding the performance of the compression scheme, it is about $10\sim 60$ minutes for normal maps of resolution $2K \times 2K$.

\subsection{Decompression for angular point query}
\label{sec:method:eval}

And as analyzed earlier, spatial prefiltering has been properly addressed using precomputation. With our compression scheme, the storage overhead of the precomputation is also greatly reduced. But it introduces a new issue: during rendering, we usually just need to find the value along individual angles or in an angular range in an NDF image. Thus, we do not need the entire 3D tensor to be decompressed, but only need to query the locations.

In this subsection, we first deal with a specific case corresponding to rendering under a point or directional light or light sampling. Given a pixel footprint, an incident direction and an outgoing direction, we would like to perform \emph{point query} instead of full extraction of the compressed data.

With our compression based on tensor decomposition, the point query can be elegantly achieved. This is because of the property of the outer product operation -- any element in the resulting rank-1 tensor is the product of corresponding elements from the resulting 1D vectors, as Fig.~\ref{fig:tc} illustrates. Therefore, given a query with an arbitrary pixel footprint and the half vector $\bh = (h_x,h_y)$ between the incident and outgoing directions, it is trivial to locate the query index $(x,y,z)$ in a clustered group $j$ using $\bx$ and $\bh$. Then we can simply extract the specific NDF value as: 
\begin{equation}
	\hat{D}(x,y,z) = \sum_{r = 1}^{R} {C_{r} \cdot \bm{X}_r(x) \cdot \bm{Y}_r(y) \cdot \bm{Z}_r(z)}.
\label{eq:decompress}
\end{equation}

Recall that when the pixel footprint is not an exact match of the precomputed ones, we perform trilinear interpolation of the NDF values from nearby precomputed pixel footprints. The is equivalent to trilinearly blending the specific NDF values towards $\bh$. Therefore, Eqn.~\ref{eq:decompress} will simply be called multiple times. Also note that our decompression for point query is as accurate as acquiring a full NDF image followed by querying. It does not introduce any further approximation error in addition to the compression itself.

\subsection{Decompression for angular range query}
\label{sec:method:eval_range}

\myfigure{tc}{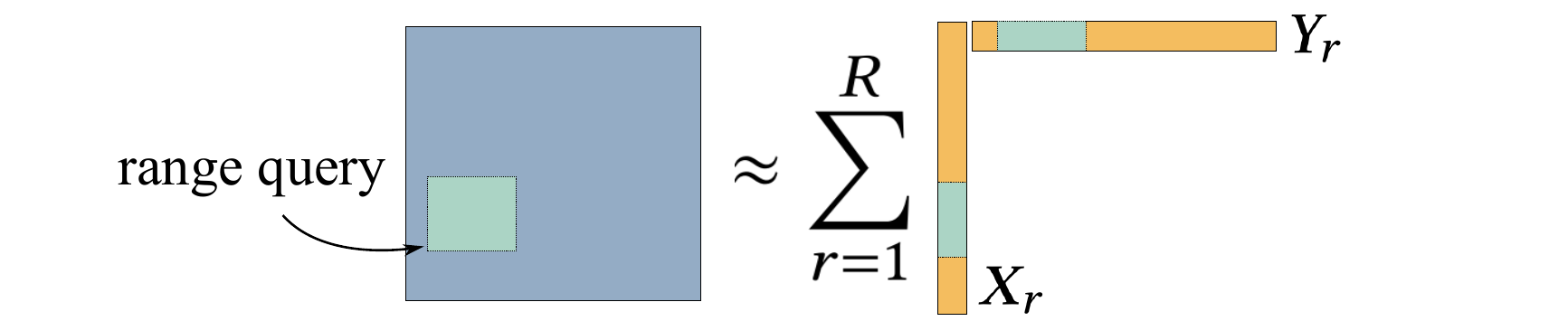}{Angular range query (green box) is performed on the compressed tensors. The average value in a rectangle can be decomposed as the product of the average values along its axis green segment on the right image).}

\myfigure{rangeAndPoint}{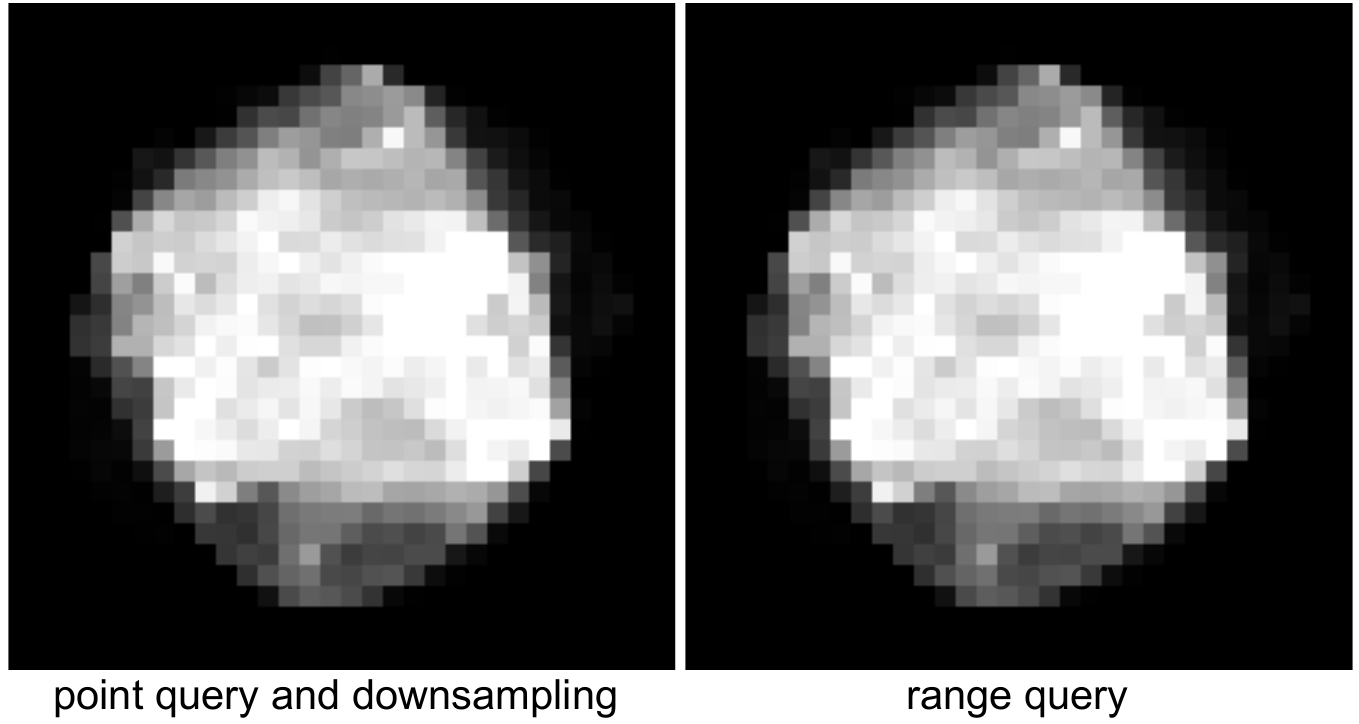}{The NDFs images are computed by (left) performing point query and downsampling and (right) performing range query directly, and they are identical.}

As motivated in Sec.~\ref{sec:motivation}, when large area lights or environment lighting are involved, angular prefiltering can greatly increase the efficiency of the rendering process. And angular prefiltering is essentially asking for the average value within a range of pixels in an NDF image. Therefore, besides angular point query, \emph{range query} is also a common operation. Moreover, as will be introduced in the next section, importance sampling also relies heavily on the range query. Hence, the efficiency of range query is crucial.

Suppose the angular range of half vectors maps to a rectangular (not necessarily square) region $[x_1, x_2]\times [y_1, y_2]$ on an NDF image block with ``depth'' $z$ in a specific clustered group $j$, a simple solution to this range query is to perform multiple point queries at different pixels using Eqn.~\ref{eq:decompress} and average them. However, this is costly and its performance scales with the size of the angular range. It seems difficult to make the performance cost of range query constant to the query size, however, with our tensor decomposition, the range query can be solved cleanly and efficiently as:
\begin{equation}
	\hat{D}([x_1, x_2], [y_1, y_2],z) = \sum_{r = 1}^{R} {C_{r} \cdot \bar{\bm{X}}_r([x_1, x_2]) \cdot \bar{\bm{Y}}_r([y_1, y_2]) \cdot \bm{Z}_r(z)}.
\label{eq:rangeDecom}
\end{equation}
where $\bar{\bm{X}}_r([x_1, x_2])$ means the average value in a segment $[x_1, x_2]$ on the 1D vector $\bm{X}$, and similarly for $\bm{Y}$.

An illustration of Eqn.~\ref{eq:rangeDecom} can be found in Fig.~\ref{fig:tc}. The key observation is that the average value in a rectangle can be decomposed as the product of the average values along its axis. Note specifically that this observation is \emph{not generally true}, but is \emph{strictly accurate in our case}. We provide proof with detailed derivation in the Appendix. We also verify its accuracy in practice in Fig.~\ref{fig:rangeAndPoint}, showing identical results generated using our range query once and our point query repeatedly.

Now the only remaining task is to quickly attain the average value in any given segment on a 1D vector. This problem is well studied with the help of the Summed Area Table (SAT)~\cite{Crow:2984:SAT} data structure. In our case, this is even easier because we only need a 1D SAT for each 1D vector. The SAT performs in linear time and exactly doubles the storage, but makes the time cost of angular range query $O(1)$ (strictly, $2$ memory look-ups) and gives exact results.

One might think of a 1D mipmap structure to perform the same task, but we would like to point out that (1) mipmaps only provide approximated range queries, (2) mipmaps also double the storage in 1D instead of introducing only $33\%$ additional storage as in 2D cases, and (3) mipmaps require $4$ memory look-ups for each query due to trilinear interpolation in 1D. Therefore, SAT is always superior.

\textbf{Practical choices.} When the angular range happen to overlap multiple image blocks, we subdivide it into smaller ones according to the boundary of the image blocks. In practice, this situation does not happen frequently. Also, throughout our paper, we report the full storage cost including the SATs.

\textbf{Summary.} In this section, we have elaborated our precomputation scheme which enables constant performance spatial prefiltering. Then we propose our compression scheme using tensor decomposition, which not only greatly reduces the storage cost, but also leads to efficient and accurate angular prefiltering, for both point query and range query. So, now we have a complete solution of constant performance spatio-angular prefiltering of glinty appearance. In the next section, we show its practical applications in different rendering tasks that were previously considered difficult. We also show how our method could combine with constant storage appearance synthesis approaches, to complete our method to an overall constant cost approach.

\section{Constant-cost glinty appearance rendering}
\label{sec:rendering}

In this section, we show several applications with our compressed NDF images. We start from efficient BRDF evaluation (Sec.~\ref{sec:evaluation}), then extend to unique applications enabled by spatio-angular prefiltering, such as accurate NDF importance sampling (Sec.~\ref{sec:sampling}), global illumination with glinty appearance (Sec.~\ref{sec:multibounce}), environment lighting prefiltering (Sec.~\ref{sec:prefiltering}) and constant storage appearance synthesis (Sec.~\ref{sec:synthesis}).

\subsection{BRDF evaluation}
\label{sec:evaluation}

With the point angular query, it's straightforward and fast to evaluate the BRDF value for arbitrary pixel footprints, incident and outgoing directions. Our approach avoids the expensive hierarchy traversal and the calculation of contribution from a large number of Gaussian flakes when the footprint is large, yielding much faster constant performance BRDF evaluation.

Since we precompute a range of pixel footprint sizes, in practice, the pixel footprint size can exceed this range. If the pixel footprint size is smaller than the precomputed size at the lowest level of the tensor mipmap, we switch to the method by Yan et al.~\shortcite{yan2016position}, which is now efficient with the small footprint size. If the pixel footprint size is larger than the largest size precomputed at the highest level, we directly clamp the size to the largest precomputed, since we stop the precomputation when we believe that the NDFs at the highest level have converged to a smooth distribution. 

\subsection{Accurate NDF importance sampling}
\label{sec:sampling}

\myfigure{sampling}{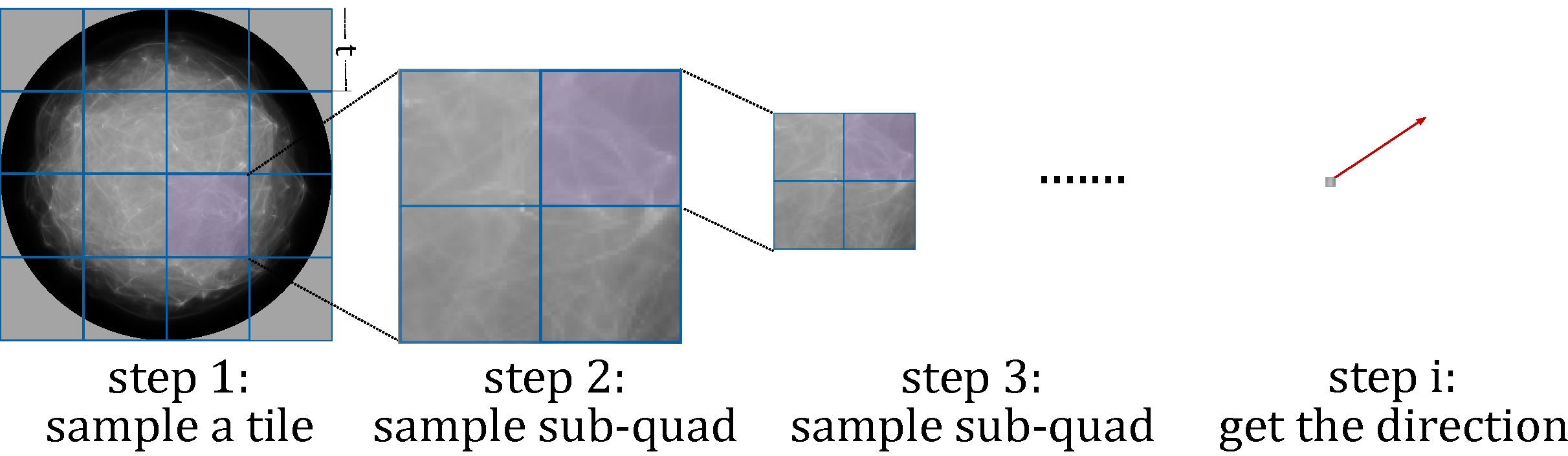}{We sample the outgoing direction hierarchically: start from choosing an image block from NDF, by sampling the averaged NDF values with CDF, and then choose sub-quad with importance sampling until reaching the pixels, and then interpolate the surrounding directions to get the final sampled direction.}

\myfigure{sampleNDF}{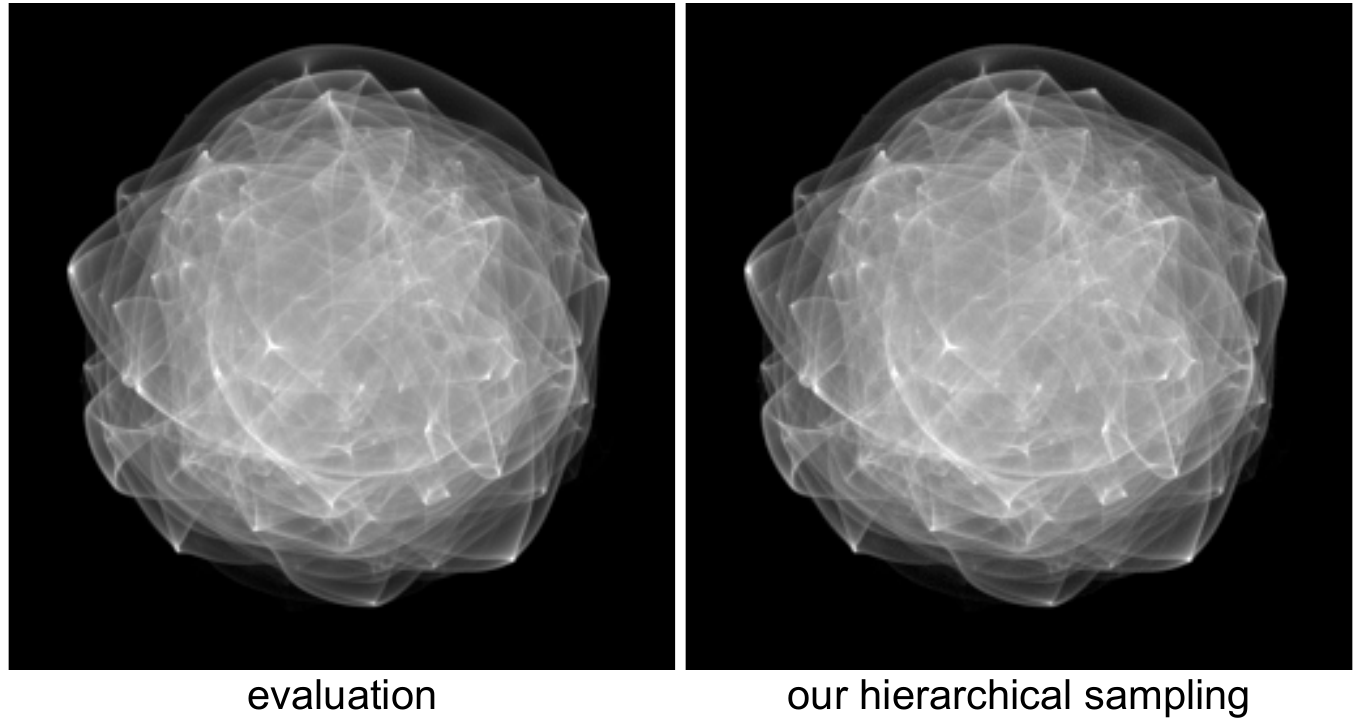}{Comparison of NDF images computed with BRDF evaluation and our hierarchical BRDF sampling via the binning method. Binning means sampling a direction and accumulate into the corresponding pixel in the NDF images, like a histogram. They are identical.}

Besides evaluation, importance sampling is another commonly used operation. Given the pixel footprint and an incident direction, importance sampling gives a sampled outgoing direction, which equivalent to a sampled half vector direction. Ideally, the half vector should be sampled strictly according to the shape of the NDF, which would reduce the variance to the minimum thus is called perfect importance sampling.

One possible way to achieve perfect importance sampling is reconstructing the entire NDF image with brute force angular point queries, computing its 2D Cumulative Distribution Function (CDF), and using the inverse sampling method to draw an outgoing direction with this CDF. However, this is very time-consuming and completely offsets the benefits from our efficient BRDF evaluation.

Thanks to our angular range query, we propose an efficient hierarchical importance sampling approach. We start from the entire NDF image, and subdivide it evenly into four quads. Recall that our angular range query answers the average value in an arbitrary range in constant time, we are able to immediately acquire the average values in these four quads. Then we use the four average values as relative probabilities, and randomly choose one quad to proceed. We subdivide this quad into four sub-quads, sample one, and continue this subdivision and sampling process iteratively until reaching the bottom level, i.e. a pixel. In this way, the probability of drawing a sample is guaranteed to be proportional to the ``brightness'' of each pixel in an NDF image. Therefore, we have achieved perfect importance sampling.

In practice, since an NDF image has already been subdivided into image blocks, we do not have to start from the topmost level of the NDF image. Instead, we first choose an image block to start, again according to their averaged NDF values as relative probabilities. Also, when we reach a pixel, we uniformly perturb the sample location inside it. The full importance sampling approach is illustrated in Fig.~\ref{fig:sampling}.

Note that since the range query is accurate, our importance sampling approach is also accurate. Therefore, the Probability Density Function (PDF) associated with the sampling method is exactly the same as the evaluated NDF value. We verify this in Fig.~\ref{fig:sampleNDF}. On the left, we show an NDF image generated using point query evaluation. On the right, we show the converged histogram of 10M sampled half vector directions (a.k.a. using the binning method). As expected, these two NDF images are exactly identical.

\textbf{Discussion.} The way we use angular range queries to do importance sampling essentially gives us an \emph{implicit} hierarchical structure on the NDF image. This results in a logarithmic performance w.r.t. the size of NDF images, which is still a constant $256\times 256$. Therefore, our importance sampling approach is still constant performance and runs efficiently in practice. We also would like to point out that building an explicit hierarchical structure on the fly for an NDF image is impractical, since this step will already take linear time, as heavy as the CDF-based sampling approach.

Unlike Yan et al.~\shortcite{yan2016position} that keeps the original normal map to facilitate importance sampling, our importance sampling no longer needs it, which further saves storage. Also note that, there are no previous method that does perfect importance sampling under wave optics. But in our framework, since we uniformly represent NDFs as precomputed images for both geometric and wave optics, no additional effort will be needed, except that we use the grayscale NDF image to conduct importance sampling.

\subsection{Global illumination with glinty appearance}
\label{sec:multibounce}

\myfigure{kettle}{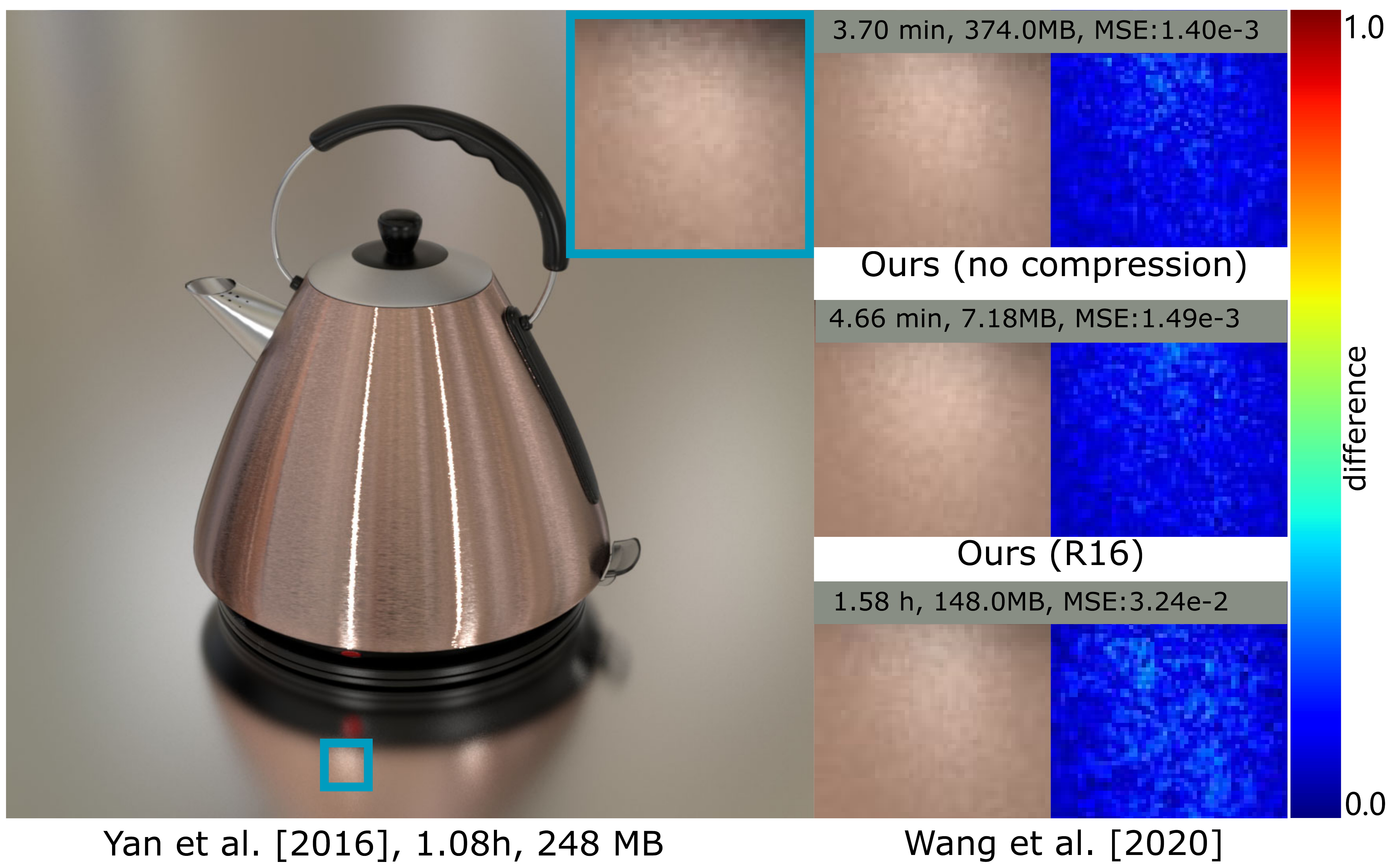}{Global illumination with glinty appearance: Comparison between our method (without compression), our method (with compression with rank 16), Wang et al.~\shortcite{Wang:2020:Example} and Yan et al.~\shortcite{yan2016position} on the Kettle Scene, considering geometric optics rendering and indirect glints. Both of our methods do not consider synthesizing. The results of these methods are identical, however our method is 14$\times$ faster. Our method with compression costs much less storage cost than our method (without compression). The picture is with $1024 \times 1024$ pixels. Normal map: brushed metal. }

Most of the previous glinty appearance rendering methods limit themselves to the direct illumination only. This is because the pixel footprints are amplified significantly as the light undergoes more bounces during the light transport. To our knowledge, Belcour et al.~\shortcite{Belcour:2017:prefiltering} deal with explicit multiple-bounce global illumination on glinty appearance. However, it only provides a way to calculate the coverages of pixel footprints at different bounces, but still has to clamp them to a small value for practical rendering, and still yields drastically increasing time cost.

With our spatial prefiltering (plus angular point query), our method automatically solves the global illumination problem, making it practical. We follow the indirect footprint computation by Wang et al.~\shortcite{Wang:2018:Glint}, using accurate footprint for the direct footprint and amply it considering the glossiness during the following bounces. In Fig.~\ref{fig:kettle}, we provide examples and discussion.

Note again that the global illumination refers to multiple bounces of light in the level of objects and scenes. It is not related to the multiple scattering of light between microfacets that leads to energy conserving BSDFs.

\subsection{Environment lighting prefiltering}
\label{sec:prefiltering}

\myfigure{prefiltering}{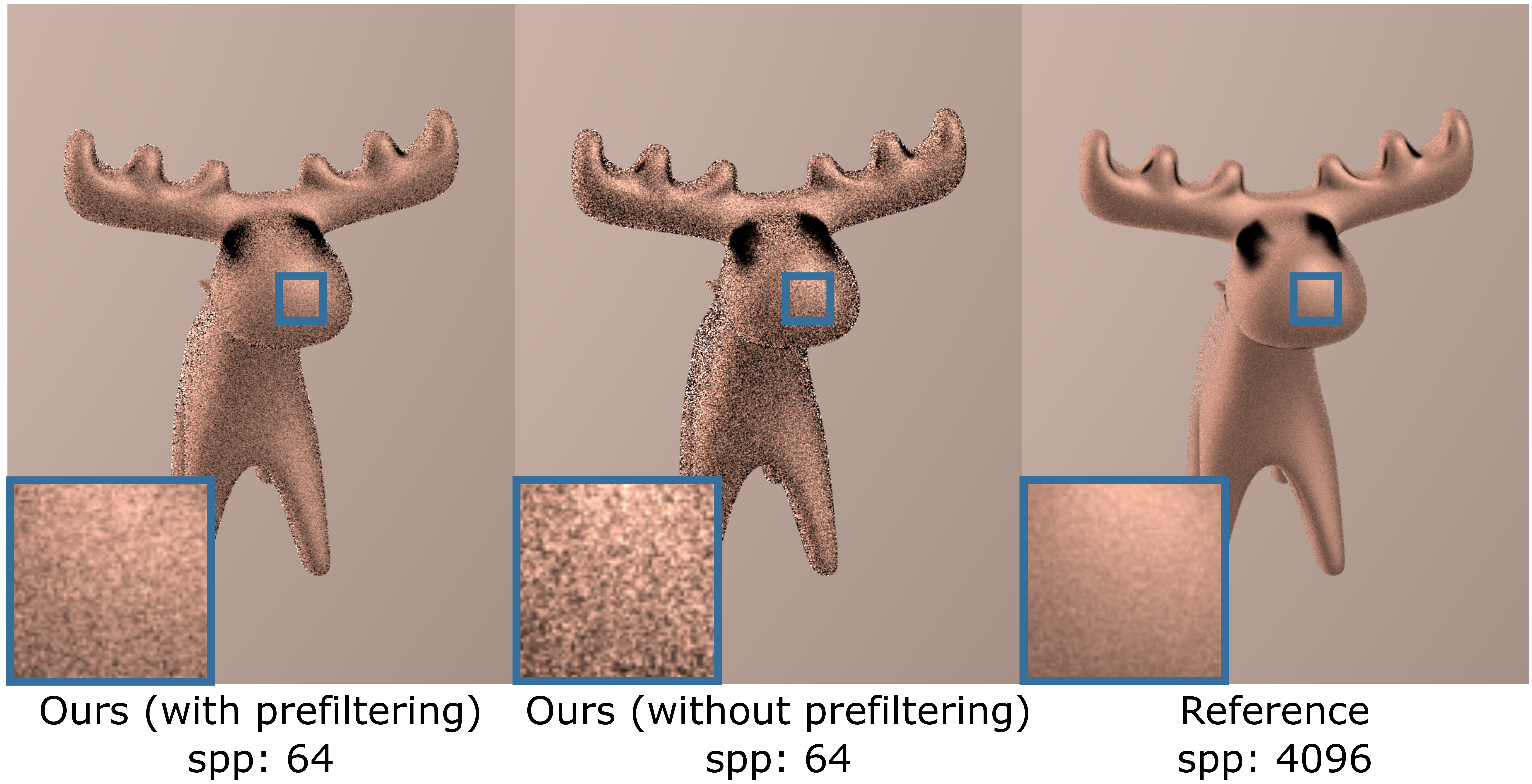} {Environment lighting prefiltering: Comparison between our method with / without prefiltering and the reference, rendered with Yan et al.~\shortcite{yan2016position}. We do not use multiple importance sampling to verify the prefiltering impact more clear. Our method (with prefiltering) produces much less noise than our method (without prefiltering). The picture is with $920 \times 1024$ pixels. Normal map: isotropic noise.}

Under distant lighting, especially defined using environment maps, rendering of the microstructures is the convolution of the incoming radiance and the BRDF, which in an integral over the directional domain. Traditionally, to compute this integral, a large number of samples are required to obtain noise-free results, even the environment map is low frequency. Given the knowledge of a low-frequency light, prefiltering could be used to reduce required samples, essentially computing blurred versions of the NDF images. Therefore, our angular prefiltering scheme can be conveniently applied in this prefiltering task.

Without loss of generality, we represent the environment map with spherical Gaussians (SGs) (\cite{Tsai:2006:SG}). Other choices are all possible as long as their frequency bandwidths can be acquired. During rendering, we first importance sample the SG-represented environment lighting to find one SG. Based on the SG's bandwidth (size or shape), we compute a corresponding angular range to query the average NDF values. 

To obtain this range, we first compute an angle $\theta$ in radians, where all directions within $\theta$ around the SG's central axis will have a value greater than a threshold $\epsilon$. This property is called SG's compact-$\epsilon$ support~\cite{Wang:2009:SG}:
\begin{equation}
\theta = \mathrm{arccos}\left(\frac{\mathrm{ln}\,\epsilon - \mathrm{ln}\,A}{\lambda}+1\right),
\end{equation}
where $A$ and $\lambda$ represent the amplitude and bandwidth of an SG, respectively. And in practice, we set $\epsilon$ as $0.3$.

Now that the incident lighting has a compact-$\epsilon$ support of $\theta$, as Fig.~\ref{fig:motivation} (right) indicates, we are able to safely ``blur'' the NDF using an SG with an approximate compact-$\epsilon$ support of $\theta/2$, which maps to a square area on the NDF image with side length
\begin{equation}
Q =  256\cdot\frac{\theta/2}{\pi}\cdot 2=256\theta/\pi
\end{equation}
where $256$ is the resolution of the NDF image.

The essential effect of our angular prefiltering is that the maximum frequencies of both the light and the BRDF are significantly reduced. This results in much lower noise level compared to naive point sampling (Fig.~\ref{fig:prefiltering}).

\subsection{Constant-storage appearance synthesis}
\label{sec:synthesis}

\myfigure{wangTiles}{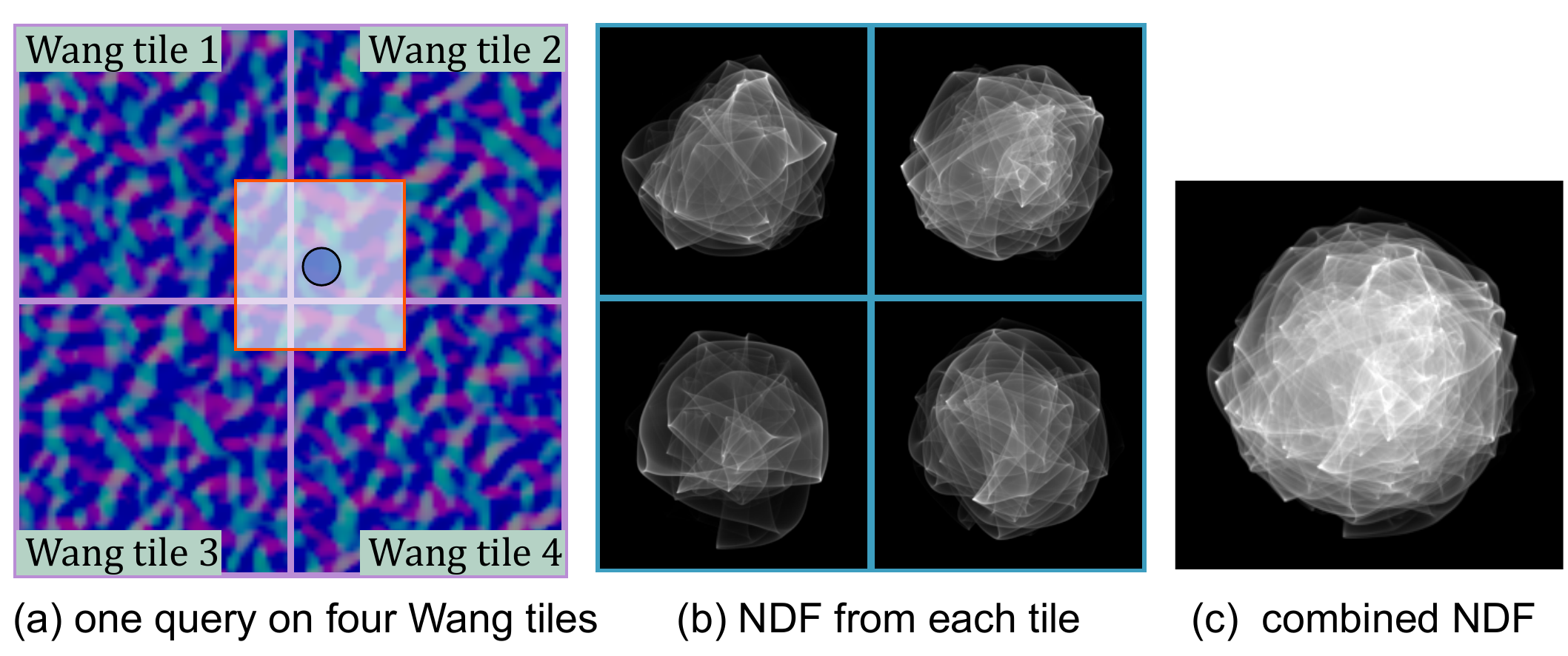}{During precomputation, we precompute the NDF images for each Wang tile, allowing queries whose centers are outside but still intersect the Wang tile. During runtime, the NDF image for a query that crosses the borders between different Wang tiles is accurately computed by combining the precomputed NDF images from all overlapping Wang tiles.}

With compressed NDF images, our method is able to achieve constant performance, even for large footprint size. \added{However, the storage cost is still expensive for high-resolution normal maps, thus, we would like to implicitly generate infinite large normal maps. Wang tiles~\cite{Cohen:2003:WangTiles} is able to synthesize large textures by repeating precomputed seamless tiles.} We propose to combine our method with Wang tiles to synthesize a high-resolution normal map from an input sample normal map for constant storage. 

\textbf{Precomputation and compression.} We generate the Wang tiles from the input normal map. Then we precompute NDF images for each Wang tile. The precomputation process is similar to the precomputation for a normal map in Section~\ref{sec:precompute}, except the sampled centers could locate outside the Wang tiles, as shown in Figure~\ref{fig:wangTiles}. In practice, we represent each normal map with $16$ Wang tiles and set the Wang tile size as $512 \times 512$. We then compress the NDF images for each Wang tile as described in Section~\ref{sec:compress} and build the SAT for compressed tensor vectors.

\myfigure{wangtilesGrid}{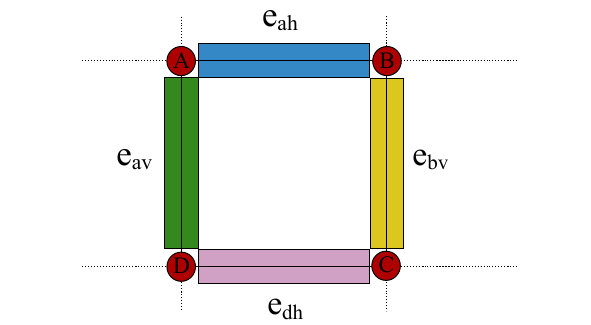}{To generate synthesized normal map on the fly without any index map, we generate two random numbers with the vertex index as seed, and then map these numbers to ``color'' via a hash table. For example, we generate two random numbers with the index of vertex $A$, and then map them to two ``color'' with the hash table. The two ``color'' are set to edges $e_{ah}$ and $e_{av}$, which are the right and bottom edges of vertex $A$. After determining the ``color'' of all the edges, we get the corresponding Wang tile.}

\myfigure{NDFwangTiles}{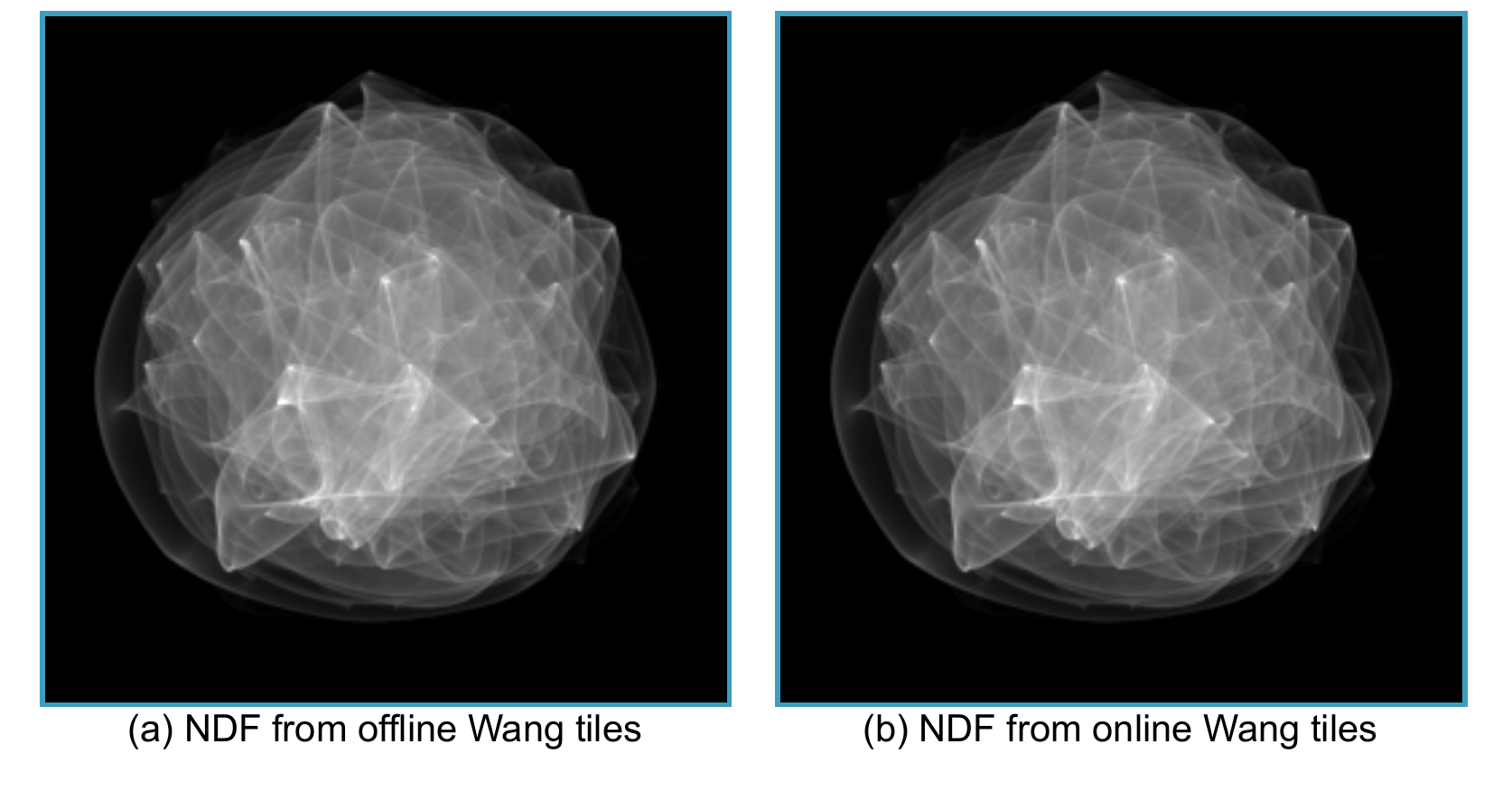}{We visualize the NDF of a normal map generated by Wang tiles offline (generate the synthesized normal map explicitly) and online (using the color hash table to generate the synthesized normal map implicitly).  }

\mycfigure{deer}{deerGan.pdf}{\added{Constant-storage appearance synthesis demonstration.} Comparison between our method with / without Wang tiles, Kuznetsov et al.~\shortcite{Kuznetsov:2019:GlintGan}, Yan et al.~\shortcite{yan2016position} and Wang et al.~\shortcite{Wang:2020:Example}, considering geometric optics rendering. Our method with Wang tiles solves the repetitive pattern issue, compared to our result without Wang tiles and Yan et al.~\shortcite{yan2016position}. Kuznetsov et al.~\shortcite{Kuznetsov:2019:GlintGan} produces results with lower quality and higher time cost, compared to our method. The picture is with $920 \times 1024$ pixels. Normal map: isotropic noise.}

\textbf{Dynamic Wang tile generation.} 
Originally, the Wang tiles method still requires a pre-generated index map to determine the locations of individual Wang tiles. Therefore, strictly speaking, it cannot generate an infinitely large normal map at the cost of constant storage. For dynamic Wang tile selection at arbitrary locations, Lagae and Dutr\'e~\shortcite{LD:2005:APODF} proposed a method that implicitly build an infinitely large grid, and associate each vertex in this grid with four random numbers. During runtime, the ``color'' of an edge in this grid is determined using its associated vertices' corresponding random numbers. We slightly improve upon Lagae and Dutr\'e~\shortcite{LD:2005:APODF} so each vertex directly controls the ``color'' of edges to its right and bottom (Figure~\ref{fig:wangtilesGrid}). For each vertex, we generate two random numbers according to a hashed value of its position / index. The two random numbers are used to determine the color of the right and bottom edges to this vertex, respectively. During rendering, given a query center and footprint size, we first find the ``colors'' of all the edges covered by the query, then we are able to identity the corresponding Wang tiles from a precomputed edge-to-tile map. All the computation is quickly performed on the fly. 

\textbf{Evaluation and sampling.}
After getting the Wang tiles covered by the footprint, we perform BRDF evaluation for each Wang tile, similar to Section~\ref{sec:method:eval} and sum all contribution of the Wang tiles, as shown in Figure~\ref{fig:wangTiles}. In Figure~\ref{fig:NDFwangTiles}, we compare the NDF images of normal maps which are generated by the implicit and explicit Wang tile algorithms, and they are identical. For importance sampling, we first choose one Wang tile from all the Wang tiles covered by the pixel footprint with equal probability. Then, we sample a direction from the chosen Wang tile, similar to Sec.~\ref{sec:sampling}. After getting the sampled direction, we computed the NDF value, which is the same as BRDF evaluation. The PDF is set the same as NDF value. In Figure~\ref{fig:deer}, we compare our method with / without Wang tiles, and other methods, to show our benefits.

\section{Results}
\label{sec:results}

 \begin{table}[htbp]
  \caption{\label{tab:texture_cost} The precomputation cost of each texture (normal map for geometric optics glints, and heightfield for wave optics glints), including NDF generation time (Pre.), compression time (Com.), total time (Tot.) and the storage cost (Stor.). Regarding the type, G represents geometric optics glints, and W represents wave optics glints. Res. is the resolution of the texture. The NDF of isotropic noise ($8K$) has three channels (RGB) while others have only one channel.}
\centering
\setlength{\tabcolsep}{3pt}
\begin{tabular}{lccccccc}
\toprule
   \multicolumn{1}{c}{Texture}
	 & \multicolumn{1}{c}{Type}
	 & \multicolumn{1}{c}{Res.}
	 & \multicolumn{1}{c}{Rank}
	 & \multicolumn{3}{c}{ Time}

	 & \multicolumn{1}{c}{ Stor.} \\	
	   
		\cmidrule(lr){5-7}
	 & 
	 & 
	 & 
	 & \multicolumn{1}{c}{ Pre.}
	 & \multicolumn{1}{c}{ Com.}
	 & \multicolumn{1}{c}{ Tot.}
	 & \\

	 & 
	 & 
	 & 
	 & min.
	 & min.
	 & min.
	 & MB \\	
		\midrule
		Metallic flakes &G &$2K^{2}$&32&42.80&60.01&102.81&106.86\\ \hline 
		Brushed metal   &G &$2K^{2}$&16&37.39&10.01&47.40&15.19\\ \hline 
		Brushed metal   &G &$1K^{2}$&16&9.56&5.36&14.92&7.18\\ \hline 
		Isotropic noise &G &$512^{2}$&16&59.71&25.32&85.03&6.87\\ \hline 
		Isotropic noise &W &$8K^{2}$&16&156.00&347.94&503.94&352.50\\ 
\bottomrule

\end{tabular}

\end{table}

 \begin{table}[]
  \caption{\label{tab:compare2micro} The rendering time (path tracing (with evaluation only) or path tracing (with sample only) of our method compared to microfacet model. All timings are measured over the Elevator scene with 1024 spp.}
\centering

\begin{tabular}{lccccccc}
\toprule
   \multicolumn{1}{c}{Method}
   & \multicolumn{1}{c}{Evaluation}
	 & \multicolumn{1}{c}{Sampling} \\	
		\midrule
		Ours        &2.30 m &6.57 m\\ \hline 
		Microfacet  &2.00 m &3.43 m\\ 
\bottomrule

\end{tabular}

\end{table}

\begin{table*}[htbp]
	\caption{\label{tab:performances} Scene settings, rendering time and memory costs for our test scenes. \#Tri. is the count of triangles in the scene. Spp. represents sample per pixel for path tracing, and different spp is used for environment map and point light in the Laptop scene. The details of input normal map or heightfield are shown in Table~\ref{tab:texture_cost}. The reference method is Yan et al.~\shortcite{yan2016position} for geometric optics rendering, and Yan et al.~\shortcite{yan2018wave} for wave optics rendering. The tile count means the repeating count of the input textures during rendering. The storage of the Laptop with Yan et al.~\shortcite{yan2018wave} is not shown, since it does not require any extra structures (flakes or acceleration structure). } 
  \centering
  \begin{tabular}{lcccccccccccc}
  \toprule
	 \multicolumn{1}{c}{Scene}
	   & \multicolumn{1}{c}{\#Tri.}
	   & \multicolumn{2}{c}{Spp. (ours)}
	   & \multicolumn{2}{c}{Spp. (ref.)}
	   & \multicolumn{1}{c}{Texture}
	   & \multicolumn{2}{c}{Tile Count}
	   & \multicolumn{2}{c}{Total time (min.)}
	   & \multicolumn{2}{c}{Storage cost (MB)} \\	
	\cmidrule(lr){3-4}
	\cmidrule(lr){5-6}
	\cmidrule(lr){8-9}
	\cmidrule(lr){10-11}
	\cmidrule(lr){12-13}
	&K&Env.&Others&Env.&Others&& Ours & Ref. &Ours&Ref.&Ours&Ref. \\
		  \midrule
		  Car      &78.1  &64  &64  &64  &64  &Metallic flakes        &1000 &1000 &5.15  &522.00 &106.86&176.00 \\ \hline
		  Kettle   &175.3 &1024&1024&1024&1024&Brushed metal($1K$)   &16   &16   &4.66  &64.80  &7.18  &44.00 \\ \hline
		  Elevator &53.5  &4096&4096&4096&4096&Brushed metal($2K$)   &1000 &1000 &64.80 &864.00 &15.19 &176.00\\ \hline
		  Laptop   &18.4  &1024&1   &1024&64  &Isotropic noise($8K$) &79   &79   &15.97 &71.17  &352.50&- \\ 

  \bottomrule
  
  \end{tabular}
  \end{table*}

\mycfigure{elevator}{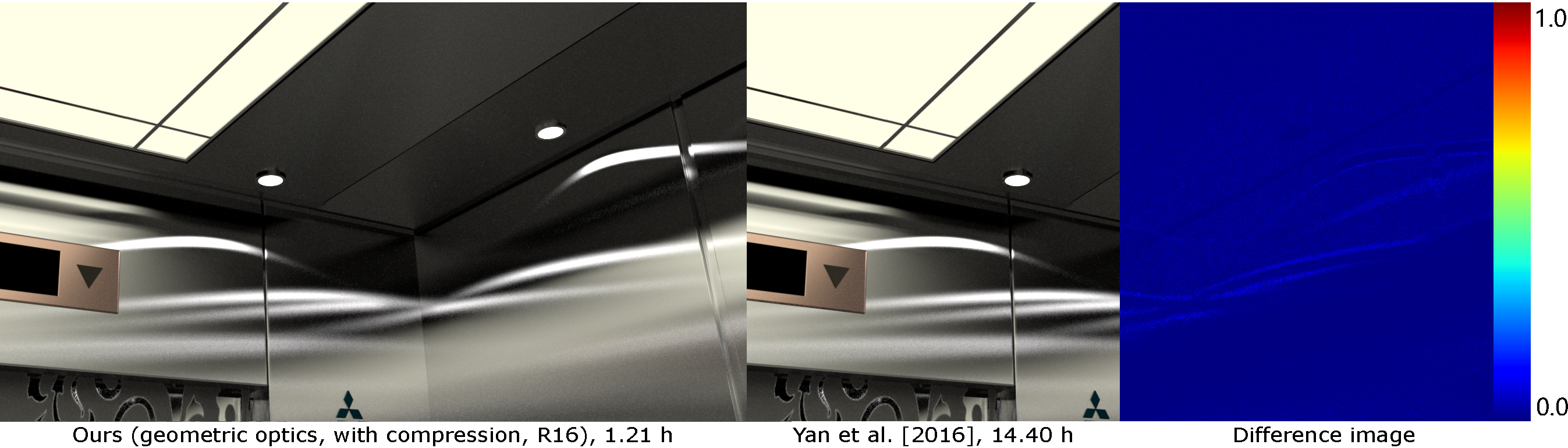}{Comparison between our method (without synthesizing) and Yan et al.~\shortcite{yan2016position} on the Elevator Scene, considering geometric optics rendering. The results of the two methods are identical, however our method is about 12$\times$ faster. R16 means using rank 16 for compression. Normal map: brushed metal.}

We have implemented our algorithm inside the Mitsuba renderer \cite{Jakob:2010:Mitsuba}. We compare against Yan et al.~\shortcite{yan2016position} for geometric optics glints validation and against Yan et al.~\shortcite{yan2018wave} for wave optics glints validation. All timings in this section are measured on a 2.39 GHz Intel i7 (40 cores) with 256 GB of main memory. Unless otherwise specified, all timings correspond to pictures with $1280 \times 720$ pixels.

\mycfigure{macbook}{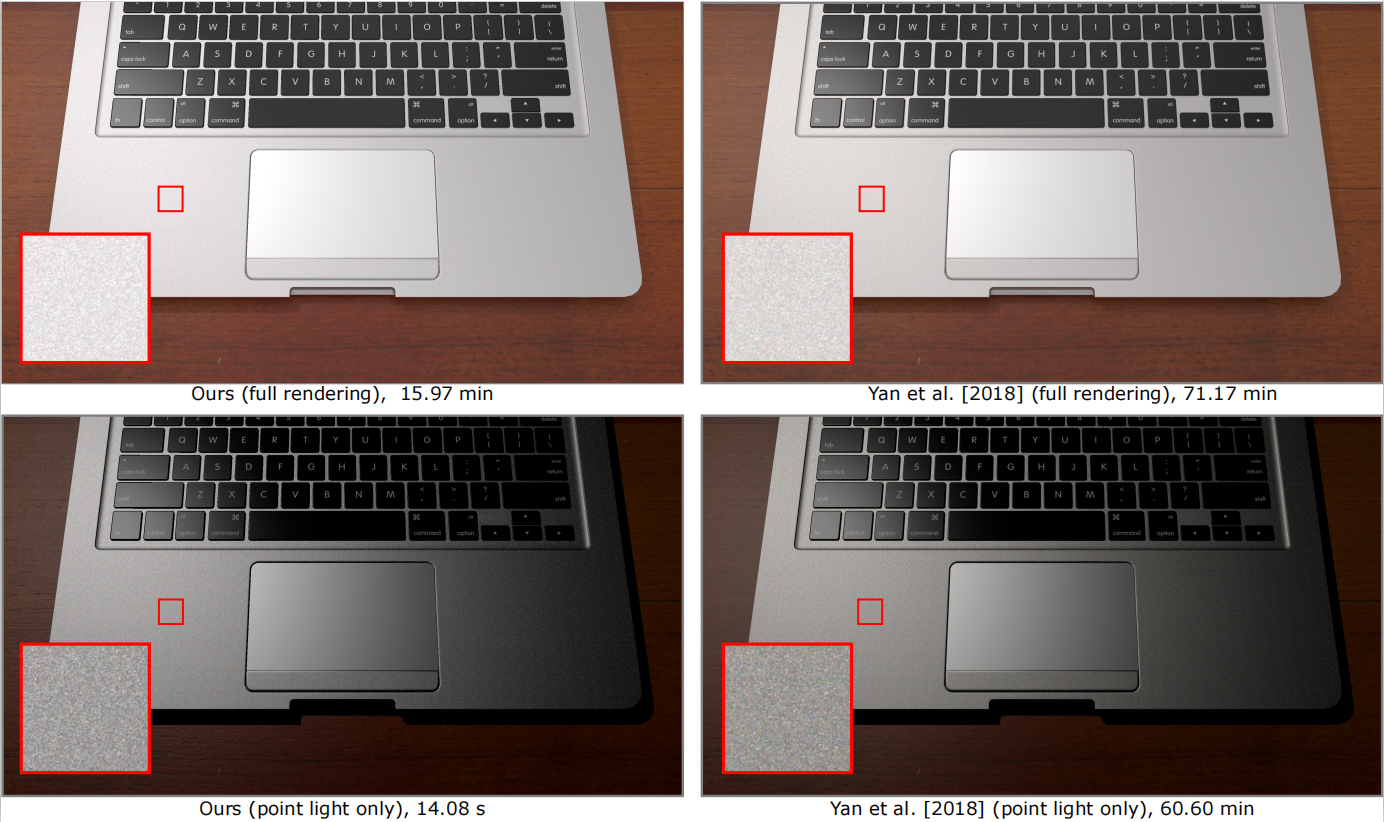}{Comparison between our method (without synthesizing) and Yan et al.~\shortcite{yan2018wave} on the Laptop Scene, considering wave optics rendering. We provided rendered results with (top row) and without (bottom row) environment map. With point light only, our method is about 260$\times$ faster than the reference. Including the environment map makes our method less impressive, about 4.5$\times$ faster, since we sample the BRDF. The picture is with $2550 \times 1426$ pixels. Heightfield: isotropic noise.}

\subsection{Performance Analysis.} 
\textbf{Precomputation Cost.} 
In Table~\ref{tab:texture_cost}, we report the precomputation cost for each normal map or heightfield, including the NDF generation time, compression time, and the storage cost. 

The compression time depends on the resolution and the NDF behaviors. For example, there are more tiles with zero-value on the brushed metal NDF image, and they are discarded directly, resulting in shorter compression time. Since rank 32 is used for metallic flake normal map, it has longer precomputation time and more storage cost. The $8K\times8K$ isotropic noise heightfield requires longer time, due to its higher resolution and more channels (RGB).

\textbf{Shading Cost.} 
In Table~\ref{tab:compare2micro}, we analysis the performance cost (BRDF evaluation and BRDF sampling) of our method during rendering, compared to mircofacet model~\cite{walter2007mmrt}. To better understand the individual cost of evaluation and sampling, we use path tracing with evaluation only or with samplings only. By comparison, the overhead of evaluation in our method is pretty low (about 15\% ), while the cost of sampling is about 2$\times$ of microfact model, due to the multiple tensor reconstruction operations.

\textbf{Rendering Cost.}
In Table~\ref{tab:performances}, we report all the scene settings, rendering time and memory costs for our method and the reference methods. The speedup of our method over the reference methods (Yan et al.~\shortcite{yan2016position} or Yan et al.~\shortcite{yan2018wave}) varies from 4$\times$ to $400\times$, depending on the footprint size, and using BRDF evaluation or sampling. For the Kettle and Elevator scenes, the high sampling rate leads to smaller footprint, which resulting in less speedup compared to the Car scene. However, high sampling rate is required for indirect illumination. For the Laptop scene, since sampling is used for environment map, it decreases the speedup. 

\subsection{Comparison with Previous Work.}
\textbf{Car scene.}
The scene in Figure~\ref{fig:teaser} shows a car with coated metallic flakes under environment lighting and a point light, with direct light only. We consider geometric optics rendering without synthesizing. The car is about $200\mathrm{cm}$ wide. The input normal map with resolution $2K \times 2K$ covers $2 \mathrm{mm} \times 2\mathrm{mm}$. We use compression rank 32 for this metallic flakes normal map, since it's discrete and requires more rank than others. Our method produces similar results to Yan et al.~\shortcite{yan2016position}, while our method is about 100$\times$ faster.

\textbf{Elevator scene.}
In Figure~\ref{fig:elevator}, we show an elevator with brushed metallic walls under several area lights, including indirect illumination. We consider geometric optics rendering without synthesizing. The wall of an elevator is about $200\mathrm{cm}$ wide. The input normal map with resolution $2K \times 2K$ covers $2 \mathrm{mm} \times 2\mathrm{mm}$. Our method is about 13$\times$ faster than Yan et al.~\shortcite{yan2016position}. Compared to Figure~\ref{fig:teaser}, the performance is less impressive due to the high sampling rate in the Elevator scene. High sampling rate results in small footprint, since the footprint of each ray is the pixel's footprint divided by the sample count, which decreases the benefit of our method. However, high sampling rate is required to remove the noise, from the indirect lighting. 
	
\textbf{Kettle scene.}
Figure~\ref{fig:kettle} illustrates a Kettle with brushed metal on the body under two small area lights and environment lighting, considering geometric optics rendering. \added{This scene is designed to show global illumination with glinty appearance.} The kettle is about $30\mathrm{cm}$ high. The input brushed metal normal map with $1K \times 1K$ resolution covers about $9\mathrm{mm} \times 9\mathrm{mm}$. Besides the direct glints, we consider the indirect glints (the glints reflected by the glossy surface). The direct glints are computed with Yan et al.~\shortcite{yan2016position} with 2.75 minutes, and the indirect glints are computed with our method with 0.68 minutes, as the footprints of direct glints are not large enough to use our method. The footprints are enlarged by the glossy surface, which yields low performance for Yan et al.~\shortcite{yan2016position}. Our method produces identical results to Yan et al.~\shortcite{yan2016position}, with about 14$\times$ speedup. We also provide the results of Wang et al.~\shortcite{Wang:2020:Example}, which is even slower than Yan et al.~\shortcite{yan2016position}.

\textbf{Laptop scene.}
The scene in Figure~\ref{fig:macbook} shows a laptop with a roughened aluminum matte finish. It is rendered using a point light and environment light, considering wave optics rendering. The laptop is about $30\mathrm{cm}$ wide. We compare our method against Yan et al.~\shortcite{yan2018wave}. In both methods, we use a tileable input texture with size $ 8K\times 8K$ with $79$ tiles. We render the results for environment map and the point light separately, which is the same as~\cite{yan2018wave}. With point light only, our method is about 260$\times$ faster than the reference, thanks to our efficient BRDF evaluation. For environment map, we perform BRDF sampling, yielding similar performance to Yan et al.~\shortcite{yan2018wave}, which slows down the speedup overall. 

\myfigure{rank_curve}{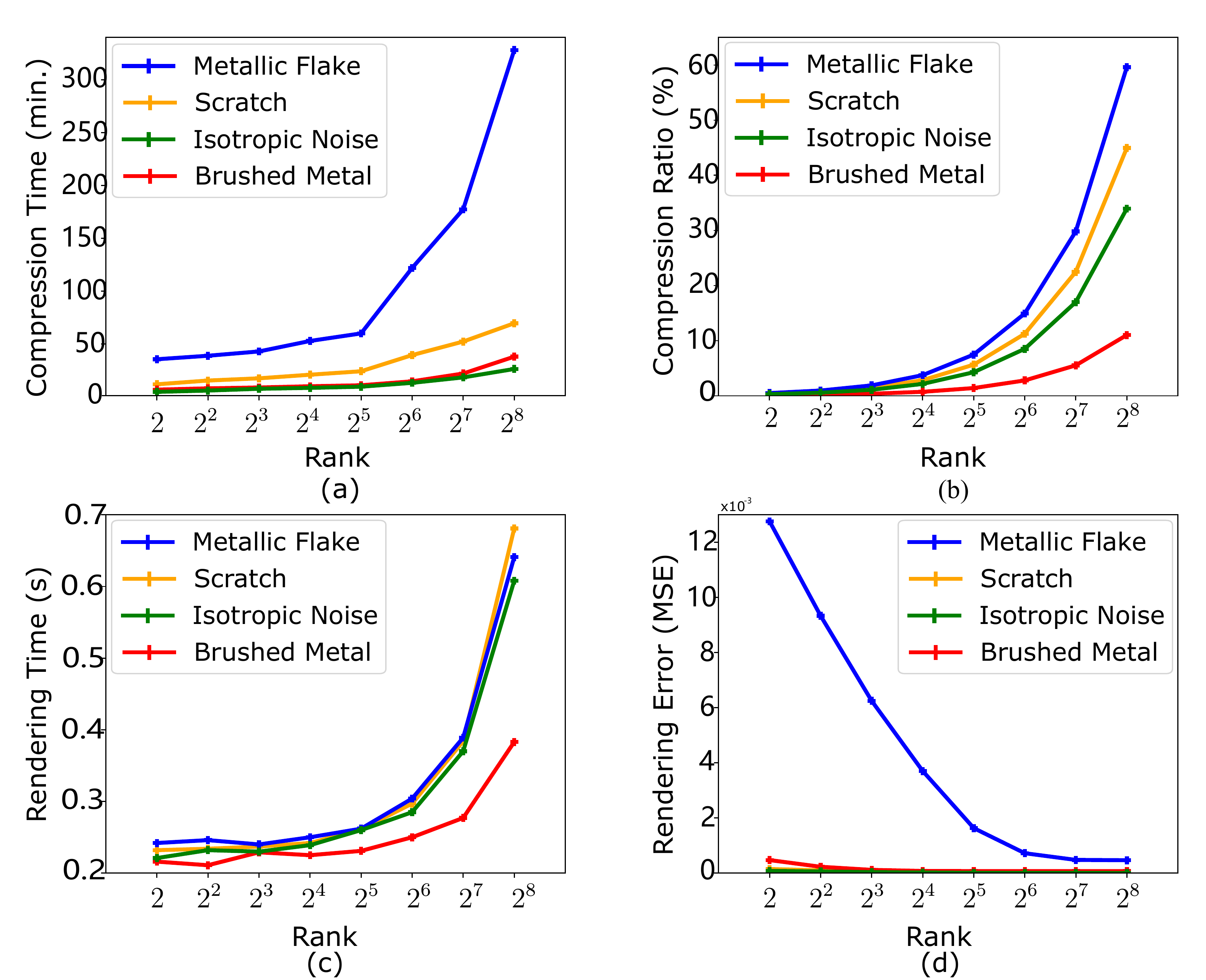}{Impact of compression rank on (a) compression time, (b) compression ratio, (c) rendering time and (d) rendered results error (MSE) on the isotropic noise normal map ($2K$). The rendering time and error are measured on the BentQuad scene with a point light.}

\myfigure{spaceship_rank}{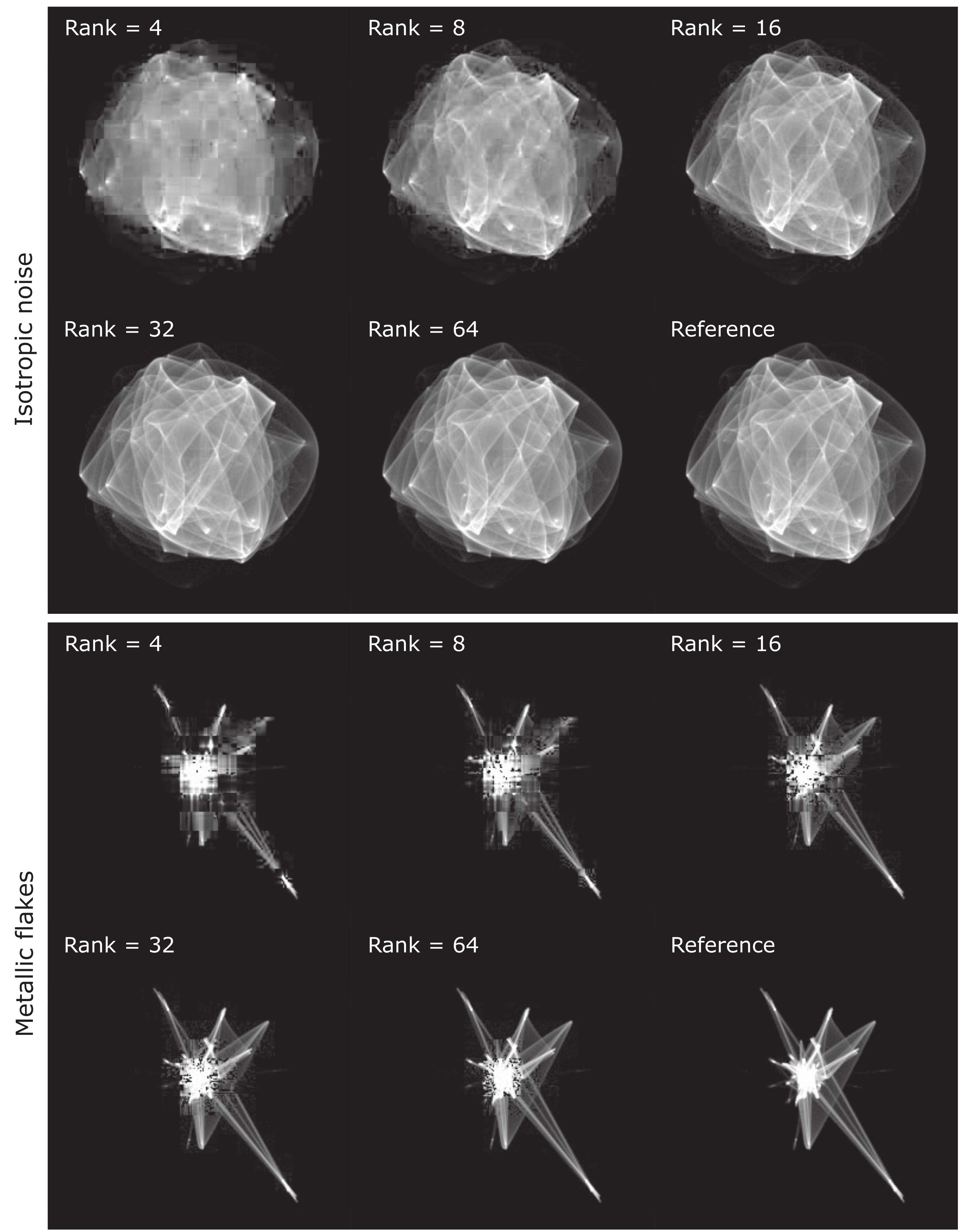}{Reconstructed NDF images from tensor decomposition with varying ranks. We set the rank as 16, balancing both the quality and the storage. Normal map: isotropic noise and metallic flakes.}

\mycfigure{renderedVaryingRank}{bentquad.pdf}{Rendered images from tensor decomposition with varying ranks on the BentQuad scene with three normal maps. We set the rank as 16 for both isotropic noise and brushed metal and set the rank as 32 for metallic flakes, balancing both the quality and the storage. }

\textbf{Deer scene.}
This scene shows a metallic deer statue under an environment lighting with geometric optics rendering. This scene is designed to show constant-storage appearance synthesis. In our method, we use 16 Wang tiles with size $512 \times 512$ from a $2K \times 2K$ normal map and the target normal map size is $100K \times 100K$. Our method with Wang tiles produces less repetitive artifacts with over little overhead. We also compare our method with Kuznetsov et al.~\shortcite{Kuznetsov:2019:GlintGan}, which introduced a Generative Adversarial Networks (GAN) model to represent NDFs. Their model only learns the NDFs at certain footprint size($256$), while our method supports multi-scale footprints. Compared to Kuznetsov et al.~\shortcite{Kuznetsov:2019:GlintGan}, our method produces better result with about $2\times$ speedup in the Figure~\ref{fig:deer}. Regarding the storage cost, our method costs $6.87 MB$, while Kuznetsov et al.~\shortcite{Kuznetsov:2019:GlintGan} only costs $1.26 MB$, as their neural network representation is more compact than ours. We also compare against Yan et al.~\shortcite{yan2016position} and Wang et al~\shortcite{Wang:2020:Example}, where Yan et al.~\shortcite{yan2016position} is 427$\times$ slower and suffers from repetitive patterns and Wang et al~\shortcite{Wang:2020:Example} is slower than Yan et al.~\shortcite{yan2016position}, but is free from the repetitive patterns. 

In Figure~\ref{fig:prefiltering}, we compare our results with / without prefiltering and the reference. With prefiltering, our method produces much less noise than without prefiltering.

\subsection{Parameter Analysis.}
Compression rank $R$ is an importance parameter in our algorithm. We show its impact on the both compression quality and compression storage. The compression quality is measured with difference between the image rendered with compressed NDF and original NDF. Figure~\ref{fig:rank_curve} shows the impact of compression rank on compression time, compression ratio, rendering time and rendered results error of our algorithm on three normal maps (isotropic noise, brushed metal and metallic flakes). As shown in the curve, larger rank yields longer compression time, more storage cost and longer rendering time but less render error. In our implementation, the rank is set as 16 for most of the test scenes, except for the metallic flake normal map, which uses 32. Different compression ratios are achieved for different normal maps, since they have different blank image blocks. For example, the NDF images of Brushed Metal normal map have the most blank image blocks thus it has the lowest compression ratio. We also show the NDF images with varying ranks in Figure~\ref{fig:spaceship_rank}. When the rank is smaller than 16, the NDFs suffer from artifacts, while with 32 or larger rank, the NDFs are almost identical to the reference (obtained without compression), thus we think 16 is a good balance considering both the storage and the quality. In Figure~\ref{fig:renderedVaryingRank}, we also show the rendered results of the BentQuad scene over varying rank on the three normal maps.

\myfigure{failure}{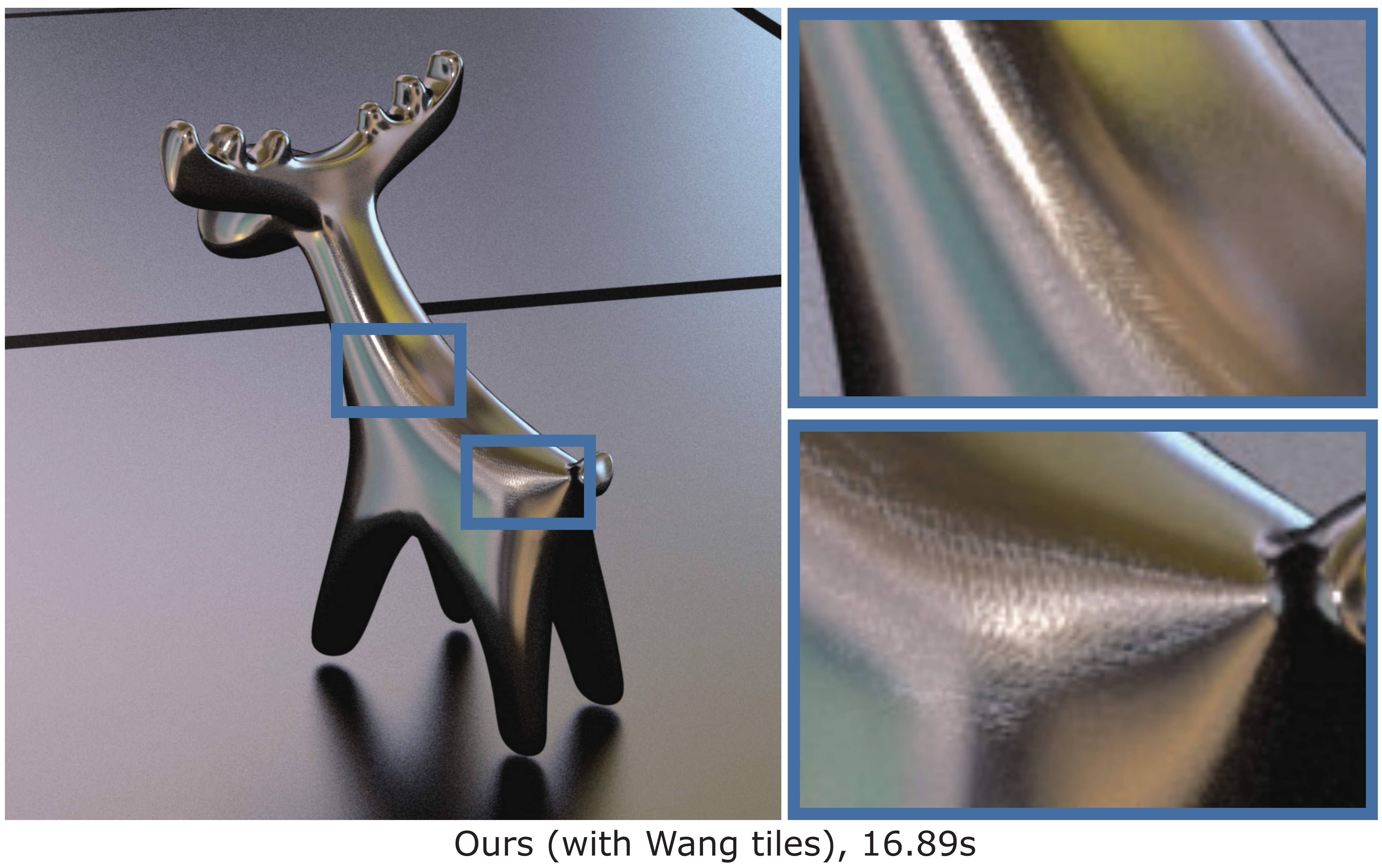}{Our method with Wang tiles suffers from the repetitive patterns for some spatial correlated textures, inheriting the drawback of Wang tiles method. The picture is with $920 \times 1024$ pixels. Normal map: scratches.}

\subsection{Discussion and limitation.}

We compare our method with Gamboa et al.~\shortcite{Gamboa:2018:GXD}, which also proposed to precompute the NDFs to achieve constant performance. They stored the NDFs with summed area table (SAT) like data structure, which has the same resolution as the input normal map, and a $9\times 32$ histogram for NDF at each texel. The NDF of an angular range could be obtained by four queries of table, however, it has very expensive storage cost, about 1.1GB for a $2K \times 2K$ input normal map. Thus, it's almost impossible to handle an $8K \times 8K$ heightfield for wave optics rendering with more than 48 GB. Besides the expensive storage, the resolution of the histogram is too low to preserve the glinty appearance of some materials (e.g. metallic flakes), as shown in our video. Thanks to our tensor decomposition to compress NDFs, our method is able to support much higher resolution ($256 \times 256$) with much less storage cost (see Table~\ref{tab:texture_cost}).

Our proposed method has several limitations. First, our method has expensive NDF generation and compression cost, although they could be reused in other scenes. Second, the compression ratio could be improved with more compact representation. Third, our method inherits the issue of Wang tiles method, suffering from visible repetitive patterns for spatial correlated normal maps, as shown in Figure~\ref{fig:failure}. Other synthesizing approach could also be coupled with our method, to solve this issue. Forth, in our prefiltering approach, we only use lights' bandwidth to limit the BRDF's bandwidth, however, the BRDF's bandwidth can also be used to limit lights' bandwidth, similar to Gamboa et al.~\shortcite{Gamboa:2018:GXD}. For example, they used a blurry environment map when the BRDF computation uses a large footprint size.

\section{Conclusion and Future Work}
\label{sec:conclusion}

We have presented a method that allows rendering of specular glints with constant performance and constant storage. By introducing compressed NDF images with angular point query and angular range query, our method is able to render glints with large footprint with constant time, and allow elegant importance sampling, prefiltering, and implicit microstructure synthesis. Eventually, our method is able to render both geometric-based glints and wave optics-based glints with constant storage and constant performance.


In the future, it would be interesting to optimize our method for real-time implementation on GPUs. Combining our method with other texture synthesis methods could also be worthwhile directions.


\bibliographystyle{ACM-Reference-Format}
\bibliography{paper}

\appendix

\section*{Appendix}
\label{sec:appendix}

\emph{Proposition:}
Suppose $\bm{D}$ is a \emph{rank-1} 2D tensor, i.e., $\bm{D}=\bm{X} \otimes \bm{Y}$. Then the average of $\bm{D}$ in range $[x_1,x_2]\times[y_1,y_2]$ equals to the product of the average of $\bm{X}$ and $\bm{Y}$ in ranges $[x_1,x_2]$ and $[y_1,y_2]$, respectively, viz.
\begin{equation}
\bar{\bm{D}}([x_1,x_2],[y_1,y_2]) = \bar{\bm{X}}([x_1,x_2]) \cdot \bar{\bm{Y}}([y_1,y_2]).\vspace{0.15in}
\end{equation}

\noindent\emph{Proof:}
According to the definition of outer product, for rank-1 tensor $D$, we immediately have 
\begin{equation}
    \bm{D}(i,j) = \bm{X}(i) \cdot \bm{Y}(j).
\end{equation}

Denoting $M=x_2-x_1+1$ and $N=y_2-y_1+1$,  

\begin{align}
    &\bar{\bm{D}}([x_1,x_2],[y_1,y_2]) \nonumber \\
    =& \frac{\sum_{i=x_1}^{x_2} \sum_{j=y_1}^{y_2} \bm{D}(i,j)}{MN} = \frac{1}{M} \sum_{i=x_1}^{x_2} \left( \frac{1}{N} \sum_{j=y_1}^{y_2} \left( \bm{X}(i) \cdot \bm{Y}(j) \right) \right) \nonumber \\
    =& \frac{1}{M} \sum_{i=x_1}^{x_2} \left( \bm{X}(i) \cdot \frac{1}{N} \sum_{j=y_1}^{y_2} \bm{Y}(j) \right) = \frac{1}{M} \sum_{i=x_1}^{x_2} \left( \bm{X}(i) \cdot \bar{\bm{Y}}([y_1,y_2]) \right) \nonumber \\
    =& \bar{\bm{Y}}([y_1,y_2]) \cdot \frac{1}{M} \sum_{i=x_1}^{x_2} \bm{X}(i) = \bar{\bm{X}}([x_1,x_2]) \cdot \bar{\bm{Y}}([y_1,y_2]).
\end{align}


Note that the derivation also trivially extends to higher dimensions.


\end{document}